\documentclass[pdflatex,sn-mathphys-ay]{sn-jnl}% Math and Physical Sciences Author Year Reference Style
\usepackage{graphicx}%
\usepackage{multirow}%
\usepackage{amsmath,amssymb,amsfonts}%
\usepackage{amsthm}%
\usepackage{mathrsfs}%
\usepackage[title]{appendix}%
\usepackage{xcolor}%
\usepackage{textcomp}%
\usepackage{manyfoot}%
\usepackage{booktabs}%
\usepackage{algorithm}%
\usepackage{algorithmicx}%
\usepackage{algpseudocode}%
\usepackage{listings}%
\usepackage{array}
\usepackage{comment}
\usepackage{soul}
\usepackage{caption}

\theoremstyle{thmstyleone}%
\theoremstyle{thmstyletwo}%

\theoremstyle{thmstylethree}%

\begin{document}

\title[VulDec: Vulnerability Detection in Binaries via Neural Decompilation]{VulDec: Vulnerability Detection in Binaries \\via Neural Decompilation}

\author*[1]{\fnm{Francesco Crescenzo} \sur{Grasso}}\email{francescocrescenzo.grasso@unina.it}
\author[1]{\fnm{Pietro} \sur{Liguori}}\email{pietro.liguori@unina.it}
\author[2]{\fnm{Roberto} \sur{Natella}}\email{roberto.natella@gssi.it}
\author[3]{\fnm{Domenico} \sur{Cotroneo}}\email{d.cotroneo@charlotte.edu}
\affil*[1]{\orgdiv{DIETI}, \orgname{Università degli Studi di Napoli Federico II}, \orgaddress{\street{Via Claudio 21}, \city{Naples}, \postcode{80125}, \state{Campania}, \country{Italy}}}
\affil[2]{\orgdiv{Computer Science}, \orgname{Gran Sasso Science Institute}, \orgaddress{\street{Viale Francesco Crispi 7}, \city{L'Aquila}, \postcode{67100}, \state{Abruzzo}, \country{Italy}}}
\affil[3]{\orgdiv{Department of Computer Science}, \orgname{University of North Carolina at Charlotte}, \orgaddress{\street{9201 University City Blvd}, \city{Charlotte}, \postcode{28223}, \state{North Carolina}, \country{USA}}}


\abstract{Automated vulnerability detection in binary executables is challenging due to the lack of source code and the limited semantic information available at the assembly level. Despite significant progress in machine-learning-based code analysis, the field still lacks methodologies and large, high-quality datasets that provide vulnerable code jointly at assembly- and source-level with precise vulnerability labels. To address this gap, we introduce VulDec, consisting of (i) a new large-scale, carefully curated dataset aligning real-world Assembly–C function pairs with detailed CWE annotations across multiple architectures, and (ii) the first end-to-end framework that leverages neural decompilation for binary vulnerability detection. 
Tuned LLM-based neural decompilers trained on our dataset achieve strong source-level recovery across architectures (Edit Similarity of approximately 56\%, BLEU-4 of approximately 43\%), while our tuned LLM-based detectors reach a macro F1-score of 0.61 on an in-distribution realistic test set (3\% vulnerability rate). Our findings provide evidence that neural decompilation offers a powerful alternative to conventional low-level binary representations, facilitating robust binary vulnerability detection.}




\maketitle

\section{Introduction}
\label{sec:introduction}

\begin{figure}
\centering
\centerline{\includegraphics[width=0.8\textwidth]{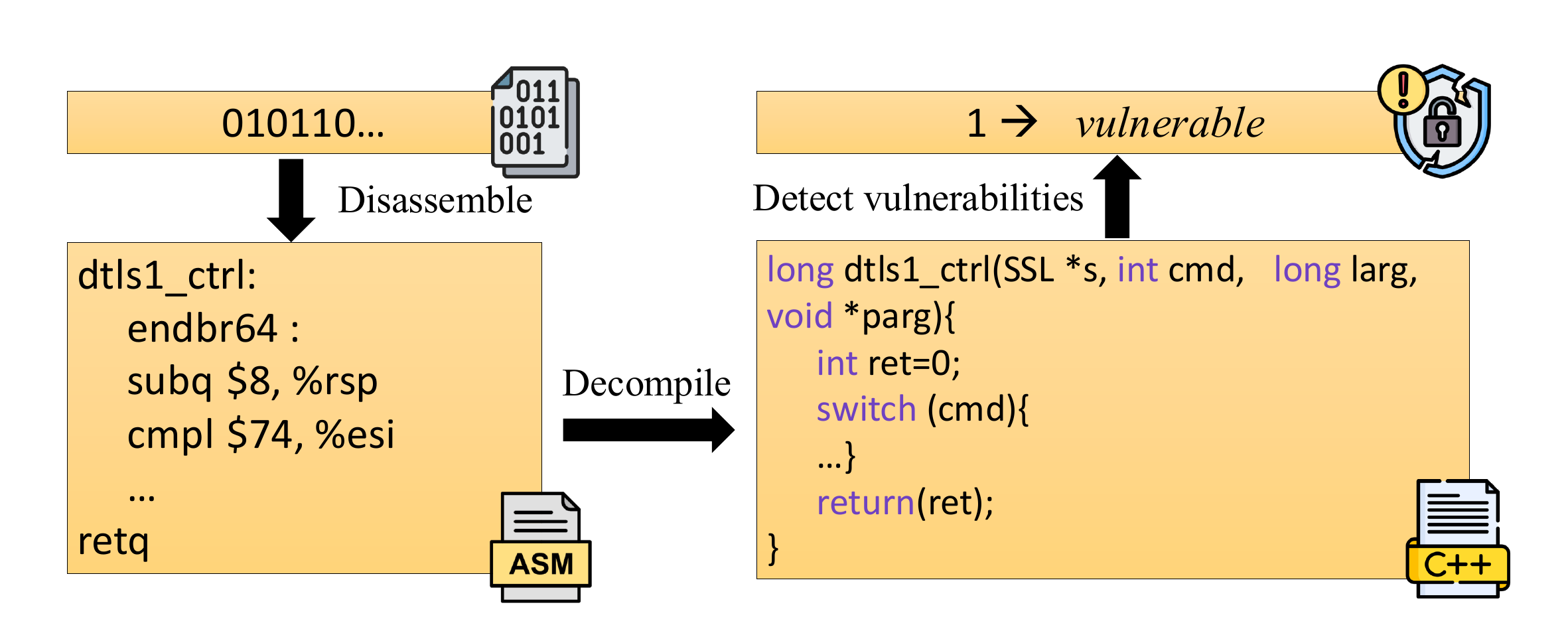}}
\caption{Illustrative example of the VulDec pipeline applied to a binary from \textit{OpenSSL}}
\label{fig:blackbox}
\end{figure}

\textit{Binary-based vulnerability detection} represents a critical challenge in automated and trustworthy software engineering, as vulnerabilities in executable files can be exploited to compromise entire systems. Unlike approaches that rely on source code \citep{arewethereyet,li2018vuldeepecker,draper,zhou2019devign}, binary-based vulnerability detection is essential in scenarios where the source code is unavailable, such as proprietary software, legacy systems, and firmware embedded in specialized hardware. As a matter of fact, many attack techniques target compiled binaries rather than source code, making binary-based analysis useful for robust system security \citep{artofwar,veribin}.

Several studies have explored binary-based vulnerability detection by leveraging different low-level representations, such as disassembled code \citep{dahl2020stack}, LLVM IR \citep{schaad2023deep,zheng2019recurrent}, and code graphs \citep{eschweiler2016discovre,grieco2016toward,pewny2015cross}. However, these representations largely preserve assembly-level structure and semantics, limiting their suitability for vulnerability detection approaches that rely on high-level abstractions rather than low-level instruction patterns.

Since vulnerability detection in source code is far more mature and supported by a well-established state of the art, binary-level detection can be approached as a standard vulnerability detection task by using \textit{decompilation}, which converts binaries into high-level code and allows existing techniques to be applied. 

As found by \citet{eff_dec}, traditional decompilers struggle to generate human-readable and well-structured code, often producing output that remains too close to raw assembly. Consequently, they fail to fully exploit the advantages of high-level program abstraction, reducing their effectiveness in automated security analysis. 

Therefore, we propose the use of \textit{neural decompilation}, a generative approach that leverages Large Language Models (LLMs) to translate binaries back into source code. The process begins by disassembling the binary into assembly code, followed by decompilation into high-level code. This methodology produces a decompiled program that closely mirrors the original source, utilizing the transformer encoder-decoder architecture to capture the semantic richness inherent in real-world code. 

Gaining access to the reconstructed source code supports a range of further applications, from malware identification to comprehensive static analysis, and, in particular, enables automated vulnerability detection. 

Along this direction, vulnerability detection on decompiled code poses non-trivial challenges. Neurally-decompiled programs, as pointed out by \citet{hosseini2022beyond}, are often incomplete, non-canonical, and lack the semantic guaranties expected by source-level analysis tools, such as precise type information and whole-program context. Most traditional methods do not natively operate at function-level granularity \citep{survey_static}; moreover, they tend to produce a high number of false positives \citep{ayewah2007evaluating}.  As a result, static vulnerability detectors tend to degrade or fail when applied to decompiled code. In this context, machine learning-based approaches, and in particular LLMs, are an appealing solution to mitigate the approximations and uncertainty introduced by decompilation, making them well-suited for vulnerability detection.

A further issue concerns existing state-of-the-art datasets: they address neural decompilation and vulnerability detection in isolation. Decompilation benchmarks \citep{angha,hosseini2022beyond,katz2018using,csmith} provide source–binary pairs without security annotations, while vulnerability datasets \citep{arewethereyet,chen2023diversevul,bigvul,megavul,Juliet,zhou2019devign} are mostly at the source-level and lack corresponding binaries. Although there exist CWE-annotated datasets that can be compiled to obtain corresponding binaries, e.g., the Juliet Test Suite C/C++ \citep{Juliet}, these resources typically rely on synthetic code. Prior work \citep{arewethereyet} has shown that synthetic samples often fail to capture real-world vulnerability patterns, thereby limiting their effectiveness in practical scenarios. Since detection operates on pseudo-C code generated by the decompiler, training both models on a shared data distribution ensures semantic consistency across pipeline stages and mitigates distribution shift between training and inference.

In light of these concerns, we propose VulDec, an end-to-end framework to detect vulnerabilities in binaries using \textit{neural decompilation} and \textit{LLM-based vulnerability detection}. Figure \ref{fig:blackbox} shows an illustrative example of our pipeline, from binary to vulnerability detection. To enable end-to-end approaches that bridge different abstraction levels, our framework includes a large-scale, carefully curated dataset that provides a direct mapping between real-world C–Assembly function pairs, complemented by detailed CWE annotations across multiple architectures.

This work aims to demonstrate that neural decompilation constitutes an effective approach for binary vulnerability detection, supporting its adoption for this purpose. 

Our contributions are as follows:
\begin{itemize}
    \item We introduce the VulDec framework, an automated LLM-based decompilation-classification pipeline  to detect vulnerabilities in binaries.
    \item  Given the lack of large-scale datasets explicitly linking low-level assembly to high-level vulnerable source code, we present the VulDec dataset \footnote{Available on \href{https://figshare.com/s/dd94f70ed12f3a11b086}{Figshare}.}, a collection of more than 59,000 (non-)vulnerable C-ASM program pairs extracted from 7,000 real-world commits, whose assembly is lifted directly from official Debian packages \citep{debian}.
    \item We perform an extensive evaluation, showing that fine-tuned LLM-based neural decompilers achieve strong source-level recovery across architectures (Edit Similarity $\approx$ 56\%, BLEU-4 $\approx$ 43\%), while our tuned LLM-based detectors reach a macro F1-score of 0.61 on an in-distribution realistic test set (3\% vulnerability rate).
\end{itemize}

The remainder of the paper is structured as follows. In Section \ref{sec:background}, we elaborate on background concepts related to the target problem. In Section \ref{sec:methodology}, we describe the construction of the dataset and the methodology adopted for vulnerability prediction through neural decompilation. In Sections \ref{sec:experimental_setup} and \ref{sec:experimental_eval}, we illustrate the experimental evaluation and analysis for the decompilation and vulnerability prediction tasks. In Section \ref{sec:lessons}, we analyze and interpret the results of the model evaluation. In Section \ref{sec:related}, we examine related work. In Section \ref{sec:threats}, we discuss the threats to validity. Section \ref{sec:conclusion} concludes the paper.
\section{Background}
\label{sec:background}

Decompilation aims to recover high-level program representations from executable code, reversing the transformations applied during compilation and linking. While the forward compilation pipeline, i.e., from high-level language to assembly, object code, and finally executable, is well defined, the reverse process is inherently ill-posed. Disassembly, which lifts binary code to assembly, is largely supported by mature tools; however, no deterministic inverse exists for decompilation, as compilation irreversibly discards high-level semantic information. This challenge primarily arises from the \textit{semantic gap} between high- and low-level code. Executable binaries lack explicit abstractions such as functions, data types, and structured control constructs, which are instead implemented through low-level jumps, calls, and register manipulations. As a result, reconstructing the original control flow and program structure from binary code is highly non-trivial. An additional drawback is the strong \textit{architectural dependence} (e.g., x86, ARM, MIPS) of low-level code. While high-level languages are largely architecture-agnostic, executable binaries are tightly coupled to the target instruction set, calling conventions, and register models. This dependency further complicates decompilation, as the same high-level construct may be realized through substantially different instruction sequences across architectures.

%The problem is further exacerbated by compiler optimizations, which may significantly alter the original code layout and semantics. Common transformations include instruction reordering, loop unrolling, or replacing arithmetic operations with equivalent low-level instruction sequences. Moreover, binaries often include platform- and operating-system–specific code that is unrelated to the program’s original logic. Consequently, traditional decompilers can only approximate the original source code, often producing representations that are syntactically correct but semantically incomplete.

\textit{Traditional decompilers} rely on static program analysis techniques to reconstruct high-level representations from binary executables. Widely used tools such as Ghidra \citep{Ghidra}, IDA Pro \citep{idapro}, and RetDec \citep{RetDec} operate on intermediate representations; P-code in Ghidra, microcode IR in IDA, and LLVM IR in RetDec. They are derived from control-flow graphs (CFGs), where basic blocks and their transitions are explicitly modeled, and may further enrich them with data-flow graphs (DFGs) and program dependence graphs (PDGs) to track variable usage and dependencies. 

However, as stated by \citet{eff_dec} and \citet{grasso}, traditional decompilers exhibit intrinsic limitations rooted in their reliance on static analysis and rule-based transformations. Their effectiveness critically depends on the availability of well-formed, \textit{fully compilable binaries}, as accurate control-flow and data-flow recovery require complete functions and closed control-flow graphs. This constraint limits their applicability in scenarios involving partial code fragments, dynamically generated code, or isolated binary snippets. Furthermore, despite traditional decompilers producing syntactically valid outputs, the reconstructed representations are often semantically approximate, \textit{assembly-like}, and verbose, reflecting low-level implementation details rather than high-level program intent. Intermediate representations are designed for deterministic structural analysis rather than semantic abstraction. Finally, the strong coupling to predefined architectures and handcrafted heuristics hinders robustness across different compiler configurations and target platforms, motivating the exploration of neural decompilation approaches that relax these assumptions by learning semantic patterns directly from data.

Previously explored by several works \citep{grasso,hosseini2022beyond,katz2018using,Tan_2024}, \textit{neural decompilation} has emerged as a promising alternative to traditional rule-based approaches by framing decompilation as a learning problem rather than a deterministic reconstruction task. Instead of explicitly modeling control-flow recovery and semantic lifting through handcrafted heuristics, neural models learn latent representations that capture recurring semantic patterns across large corpora of compiled programs.
This paradigm enables training on paired sequences of low-level and high-level code, effectively casting decompilation as a \textit{translation task}. Such a data-driven approach facilitates the implicit modeling of semantic patterns that are difficult to encode explicitly, allowing neural decompilers to recover representations that are generally more \textit{C-like}, structured, and readable than those produced by traditional tools. A key advantage lies in their increased \textit{flexibility} with respect to input completeness and variability, as neural models can operate on code fragments or isolated functions, making them applicable even when fully compilable binaries are unavailable.

Despite these advantages, neural decompilation also presents notable challenges. Code generation models inherently carry the risk of not consistently providing syntactically correct or relevant code. The performance of neural models heavily depends on the availability and quality of large-scale paired datasets of binaries and source code, whose construction is non-trivial and may introduce bias toward specific compilers, optimization levels, or programming styles. Additionally, neural decompilers may struggle with generalization beyond the training distribution and often lack the interpretability and deterministic guarantees provided by traditional analysis-based tools. As a result, neural decompilation should be viewed not as a replacement but as a complementary approach to traditional decompilation, particularly in contexts requiring scalable and semantics-aware analysis.

\textit{Vulnerability detection} is a well-established problem in software security, aiming to identify programming errors and security flaws that may lead to exploitable behaviors. Existing approaches can be broadly categorized into source-level and binary-level vulnerability detection. Source-level techniques \citep{arewethereyet,chen2023diversevul,COTRONEO2025107572,li2018vuldeepecker,draper,flawfinder,zhou2019devign} operate on high-level code representations and benefit from explicit program semantics, structured control flow, and rich contextual information, enabling precise analysis through static analysis, dynamic analysis, symbolic execution, or learning-based methods. ML-based approaches \citep{dahl2020stack,eschweiler2016discovre,grieco2016toward,  schaad2023deep, pewny2015cross, zheng2019recurrent} are particularly suitable for large or diverse code fragments, as they can capture recurring vulnerability patterns without requiring extensive handcrafted rules. In contrast, binary-level vulnerability detection operates directly on executable code, where the lack of high-level abstractions and semantic information significantly complicates the analysis. 

Decompilation plays a crucial role in enabling \textit{binary-level vulnerability detection} by recovering higher-level program representations from executables, thus allowing the application of mature source-level vulnerability detection techniques in binary-only settings. In this context, neural decompilation, by producing more structured and semantically expressive representations, provides a promising mechanism for transferring source-level security analyzes to binaries. However, despite this potential, decompilation has largely remained a support tool for code readability and manual reverse engineering \citep{hosseini2022beyond}, rather than being systematically integrated into automated security analysis pipelines. Consequently, existing approaches often fail to fully exploit recovered semantic information for security-driven tasks, motivating the development of methods that tightly couple neural decompilation with vulnerability detection to enable scalable and effective binary-level analysis.

Existing datasets typically support neural decompilation and vulnerability detection in isolation. Decompilation datasets \citep{angha,hosseini2022beyond,katz2018using,csmith} focus on paired source–binary representations but lack security annotations, whereas vulnerability detection datasets \citep{arewethereyet,chen2023diversevul,bigvul,megavul,Juliet,draper,zhou2019devign} are primarily defined at the source level and do not include corresponding binaries. Many widely used vulnerability datasets are synthetic \citep{Juliet}, cover a limited set of vulnerability patterns \citep{draper}, or do not strictly follow standardized taxonomies such as CWE \citep{arewethereyet,zhou2019devign}. Moreover, they rarely ensure full compilability \citep{arewethereyet,chen2023diversevul,bigvul,zhou2019devign}, limiting their applicability in studying how source-level vulnerabilities manifest at the binary level. As a consequence, unified datasets that jointly capture source code, binaries, and vulnerability annotations remain largely unavailable, hindering the development of end-to-end approaches that integrate decompilation and vulnerability detection.

\section{Methodology}
\label{sec:methodology}

This section introduces the proposed VulDec framework for binary vulnerability detection and the construction of the VulDec dataset used to support its training and evaluation.

\subsection{Dataset Construction}

\begin{figure*}[!htbp]
\centering
\centerline{\includegraphics[width=\textwidth]{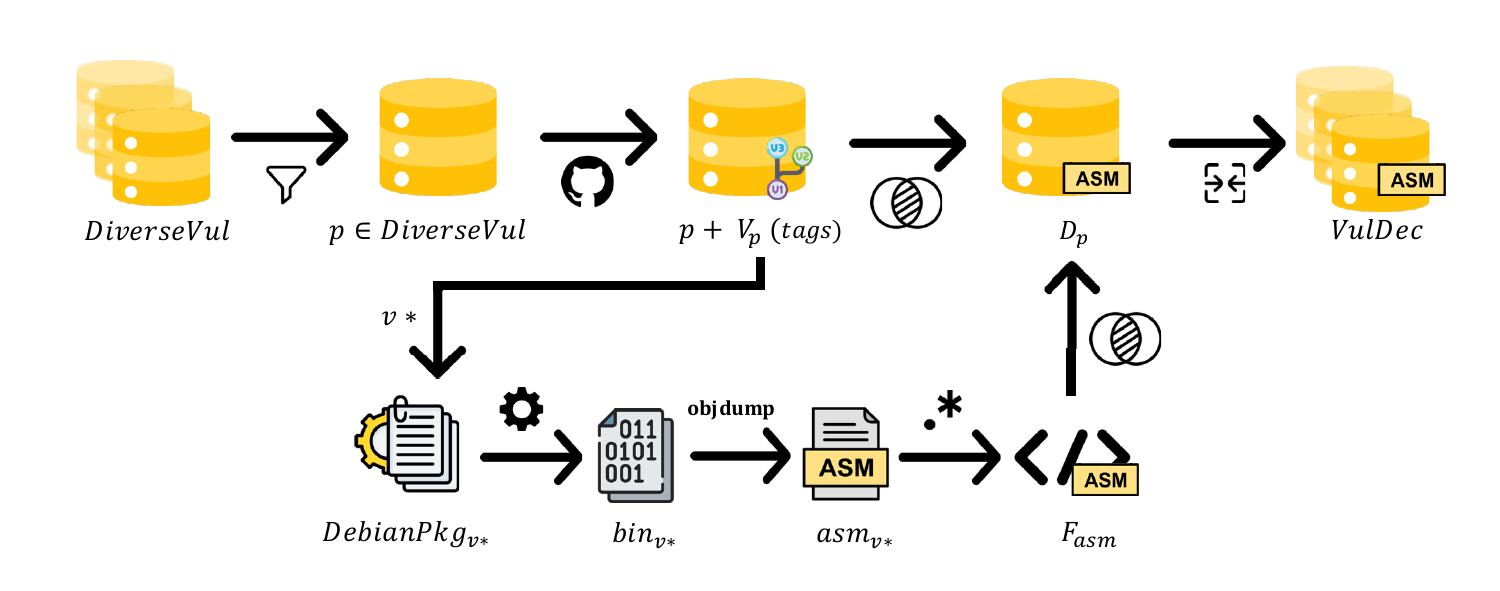}}
\caption{Automated compilation, cross-architecture assembly generation and source-level alignment}
\label{fig:Dataset Building}
\end{figure*}

The effectiveness of neural decompilation for vulnerability detection critically depends on the availability of a high-quality dataset that provides aligned source-level and binary-level representations, along with reliable vulnerability annotations. 
To this end, we constructed VulDec, a corpus of C–Assembly function pairs enriched with standardized vulnerability labels, with a particular emphasis on reproducibility and coverage of real-world software. Figure \ref{fig:Dataset Building} shows the steps leading to the construction of the dataset. 

We build upon the \textit{DiverseVul} dataset by \citet{chen2023diversevul}, a large-scale collection of vulnerable and non-vulnerable C/C++ functions mined from real-world open-source projects and annotated via vulnerability-fixing commits. Although DiverseVul does not provide directly compilable source projects, it offers a uniquely large and diverse set of source-level functions with reliable, standardized vulnerability annotations, overcoming the limitations related to the synthetic datasets \citep{arewethereyet} and those reported by \citet{jain-centric}. With approximately 19,000 vulnerable and over 330,000 non-vulnerable functions spanning 150 CWE categories, DiverseVul provides the scale required to support data-intensive neural models, motivating its selection as the foundation of our dataset. To obtain the binary representations required for neural decompilation, we leverage the availability of \textit{Debian source packages} \citep{debian}, which provide curated, reproducible, and compilable versions of many widely used open-source projects. 

\begin{algorithm}
\caption{Dataset Construction from DiverseVul}
\label{alg:construction}
\begin{algorithmic}[1]

\State $VulDec \gets \emptyset$

\ForAll{$p \in \textit{DiverseVul}$}

    \State \texttt{git clone}$(p_{url})$
    \State $V_p \gets$ \texttt{git tag}$(p)$
    \State $S_p \gets \{ s = (f_c, t, c_{id}, \text{cwe}) \mid s \in p \}$

    \State $A_p \gets \{ v \in V_p \mid \exists\ \text{DebianPkg}_v \}$

    \If{$A_p \neq \emptyset$}
    
        \State $v^* \gets \arg\max_{v \in A_p} |S_p(v)|$
        
        \State $bin_{v^*} \gets \text{build}(\text{DebianPkg}_{v^*})$
        \State $asm_{v^*} \gets \text{disassemble}(bin_{v^*})$
        \State $f_{asm} \gets \text{f\_extract}(asm_{v^*})$

        \ForAll{$s \in S_p$}
            \If{$\text{name}(f_{asm}) = \text{name}(s.f_c)$}
                \State $D_p \gets D_p \cup 
                \{(s.f_c,\; f_{asm},\; s.t,\; s.c_{id},\; s.cwe)\}$
            \EndIf
        \EndFor

        \State $VulDec \gets VulDec \cup D_p$

    \EndIf

\EndFor

\State \Return $VulDec$

\end{algorithmic}
\end{algorithm}

The process, formalized in the Algorithm \ref{alg:construction}, begins with the systematic selection of software projects from the DiverseVul. For each project $p$, a local copy is instantiated via git clone to enable deep temporal analysis of the codebase. The alignment phase leverages the project's versioning history. We extract the set of all available release tags ($V_p$) to map the specific vulnerability samples ($S_p$), comprising the source function $f_c$, the ground truth label $t$, the commit identifier $c_{id}$, and the CWE category, onto stable software versions. To ensure the reproducibility and reliability of the compilation environment, we define $A_p$ as the subset of versions for which a corresponding Debian package is available. A critical step in our methodology is the selection of the optimal version $v^*$. This is achieved through a greedy maximization strategy:$$v^* \leftarrow \text{argmax}_{v \in A_p} |S_p(v)|,$$

\noindent
where $|S_p(v)|$ is the number of samples obtained from version $v$. This choice is made to ensure that we capture the highest possible density of vulnerability samples for each project if, and only if, $A_p \neq \emptyset$, i.e, at least one version of the project is supported by the Debian build system; otherwise, we skip to the next project. Once the optimal package $\text{DebianPkg}_{v^*}$ is identified, it is compiled into a binary executable ($bin_{v^*}$). We then use a disassembler utility to obtain the assembly representation ($asm_{v^*}$) of the binary code, as well as the names (symbols) and addresses of functions in the code. From this output, the function boundaries are identified and extracted ($f_{asm}$). The dataset for a project ($D_p$) is then populated through a name-based intersection: a sample is included only if there is an equality between the symbol name in the assembly $f_{asm}$ and the function name in the source code $f_c$. 

In rare instances involving homonymous functions, a manual verification was conducted by the authors to resolve potential mapping ambiguities. We took measures to mitigate any systematic errors arising from such edge cases, ensuring their impact is negligible; a detailed discussion on these factors is provided in Section \ref{sec:threats}. 

In order to reduce syntactic variability and facilitate downstream learning tasks, both C and assembly representations undergo a normalization phase aimed at abstracting literal strings using a placeholder token $STR$, removing non-semantic artifacts ($//$ and $/*...*/$ comments, pre-processor and call frame information directives, irregular spacing) while preserving the functional structure of the code.

This rigorous automated process yields the global dataset, a collection of high-fidelity, cross-domain pairs ready for downstream tasks.

\subsection{Proposed Framework}

\begin{figure*}[!htbp]
\centering
\centerline{\includegraphics[width=0.95\textwidth]{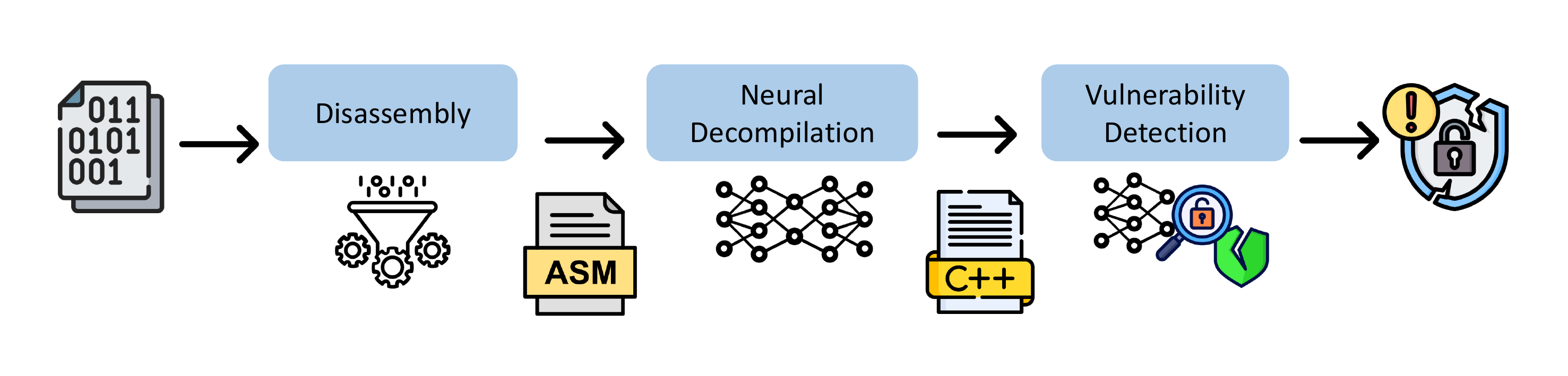}}
\caption{VulDec workflow: binary-level function analysis and vulnerability discovery}
\label{fig:Methodology}
\end{figure*}

This work introduces VulDec, an end-to-end framework for vulnerability detection directly from binary executables. Given a compiled binary, VulDec first extracts its assembly representation using a static, architecture-aware disassembler. The resulting assembly code is then processed by a neural decompilation module to reconstruct a high-level approximation of the original source code. Finally, a neural vulnerability detection module analyzes the recovered source representation to determine whether the function contains security vulnerabilities. Figure \ref{fig:Methodology} illustrates the overall pipeline and the intermediate representations produced at each stage. 

\begin{figure*}[!htbp]
\centering
\centerline{\includegraphics[width=0.9\textwidth]{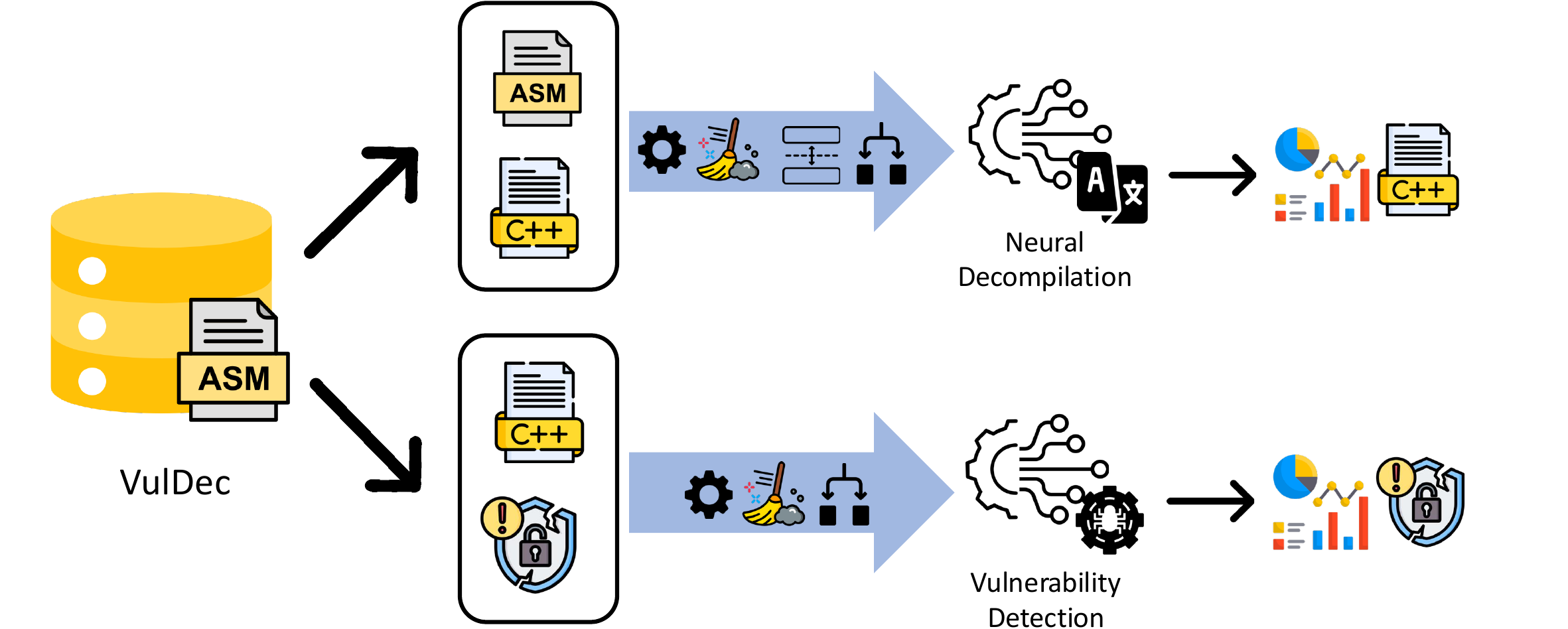}}
\caption{VulDec training and testing workflow: neural decompilation and vulnerability detection. Note: integrated data sources (Section \ref{sec:experimental_setup}) are used to enhance detection diversity}
\label{fig:Training}
\end{figure*}

As shown in Figure \ref{fig:Training}, the framework is trained through two distinct yet complementary learning tasks. Neural decompilation is formulated as a \textit{sequence-to-sequence} translation task, in which pre-trained large language models are fine-tuned to map sequences of assembly instructions to semantically equivalent high-level source code written in C or C++. The objective is to learn a conditional transformation from low-level binary representations to structured, human-readable code. All samples are defined at the function level, with variable-length instruction and token sequences, and the decompilation process operates in a controlled setting that excludes compiler optimizations and code obfuscation.

Vulnerability detection, in contrast, is formulated as a binary classification task. Here, pretrained large language models are fine-tuned to analyze the decompiled C/C++ code and predict whether a function is vulnerable (1) or non-vulnerable (0). The detection model is trained on C/C++ source code and is applied at inference to decompiled pseudo-C/C++, reflecting realistic deployment conditions.
By integrating disassembly, neural decompilation, and automated vulnerability analysis, VulDec enables vulnerability assessment in the absence of the original source code. 

\section{Experimental Setup}
\label{sec:experimental_setup}

All experiments were conducted on a workstation equipped with an \textit{NVIDIA GTX} 1660 GPU with 6 GB of VRAM and 32 GB of system RAM. For models exceeding 1 billion parameters, an \textit{NVIDIA T4} GPU with 16 GB of VRAM was additionally utilized, along with \textit{Low-Rank Adaptation} (LoRA) to mitigate the computational load. This hardware setup, combined with memory-efficient adaptation techniques, proved sufficient to support training and inference across all experimental phases, enabling repeated evaluation cycles and consistent comparisons of model configurations and preprocessing choices.

\subsection{Dataset Overview}
VulDec is a real-world C/ASM dataset for function-level vulnerability analysis across x86 and ARM architectures. 
We adopted two widely deployed ISAs (Instruction Set Architectures), representative of both server and embedded/IoT domains. 
These architectures follow fundamentally different design philosophies (CISC and RISC), allowing us to capture substantial variability at the assembly level and evaluate the generality of the proposed approach across clearly different low-level representations. The automated compilation of the binary artifacts was performed using the standard Debian toolchain (via \texttt{apt-get --build source}). 
We targeted the source packages through the distribution's build utilities, employing the GNU Compiler Collection (GCC) version 12.2.0 compiler suite with the \texttt{-O2} optimization level to ensure a consistent and reproducible production environment. 
The assembly code is obtained from binary executables using \texttt{llvm-objdump} \citep{LLVMobj}, a static, architecture-aware disassembler operating at the instruction level.

\begin{table}[htbp]
\caption{VulDec statistics on projects, CWEs, functions, and code metrics}
%\begin{center}
\small
%\resizebox{0.7\columnwidth}{!}{%
\begin{tabular}{l|cc||l|cc}
\toprule
\multicolumn{1}{c}{\textbf{Dataset}} & \textbf{ARM} & \textbf{x86} & \multicolumn{1}{c}{\textbf{Dataset}} & \textbf{ARM} & \textbf{x86} \\
\midrule
\# \textit{functions} & $28,426$ & $31,340$ & \# \textit{projects} & $24$ & $24$\\
\# \textit{vulnerable func.} & $1,879$ & $1,949$ & \# \textit{ CWE Types} & 80 & 74  \\
\% \textit{vulnerable rate} & $6.6\%$ & $6.2\%$ & \textit{LOC (median)} C/C++ & $18$ & $18$\\
\# \textit{common func.} & $23,163$ & - & \textit{LOC (median)} ASM  & $56$ & $49$\\
\bottomrule
\end{tabular}
\label{tab:Dataset}
\end{table}

Table \ref{tab:Dataset} shows some relevant statistics of VulDec. The ARM and x86 subsets share $23,163$ common functions, representing functions present in both architectures. The dataset contains no duplicate entries or missing values, ensuring data consistency. 
Vulnerability prevalence is low and reflects real-world conditions. Both architecture-specific subsets exhibit a vulnerability rate of approximately 6–7\%. This is slightly higher than the original DiverseVul rate of 5.7\% \citep{chen2023diversevul}, leading to a marginally more balanced yet still realistic class distribution. 

VulDec covers a diverse set of widely used open-source projects, including Linux, OpenSSL, QEMU, FFmpeg, and Git, with substantial variability in project size. Across both architectures, the most prevalent vulnerability classes are consistently CWE-703, CWE-119, and CWE-20, indicating that improper control flow validation, memory corruption vulnerabilities, and input validation issues dominate in both ARM and x86 binaries. Detailed dataset statistics and per-project function counts for both architectures are reported in Appendix 
\ref{app:vuldec}. 

Moreover, it is worth noting that functions expressed in high-level code exhibit a median number of lines of code (LOC) that is less than half of their assembly counterparts, which is expected given the higher abstraction level of source code.

\subsection{Decompilation Setup}
For the neural decompilation phase, transformer-based models were fine-tuned to translate assembly code into C/C++ source code. To assess the suitability of different neural architectures for translation, we adopted several representative models from diverse design paradigms. All models are pre-trained on large-scale code corpora and are subsequently fine-tuned on VulDec for the neural decompilation task.
\begin{itemize}
    \item \textit{Qwen2.5-Coder 3B} \citep{hui2024qwen25codertechnicalreport}, a decoder-only large language model optimized specifically for code, included to evaluate the capabilities of a highly capable, dense, instruction-tuned model trained on extensive multilingual code repositories.
    
    \item \textit{DeepSeek-Coder-Instruct 1.3B} \citep{guo2024deepseekcoderlargelanguagemodel} is selected to assess the performance of a instruction-tuned decoder-only model that leverages repository-level code completion and a mix of project-level context during its pre-training phase.
    
    \item \textit{CodeT5+ 220M} \citep{wang2023codet5} is employed as a reference encoder–decoder model, leveraging span-denoising pre-training and a hybrid architecture that has demonstrated strong performance in both code understanding and generation tasks.
    
    \item \textit{CodeBERT 125M} \citep{feng2020codebert}, an encoder-only model based on RoBERTa, is included to evaluate the effectiveness of rich contextual representations during the encoding of assembly instructions. In our approach, it is coupled with a seq-to-seq decoder, forming a full encoder–decoder architecture.
    
    \item \textit{CodeGPT 124M} \citep{radford2019gpt2} and \textit{CodeParrot 110M} \citep{face2023codeparrot}, both decoder-only autoregressive models, are adopted to assess the flexibility of generation-only architectures when applied to a translation task not explicitly targeted during pre-training.
    
    \item \textit{PLBART 140M} \citep{plbart}, an encoder–decoder model derived from BART and trained using multi-task denoising objectives on code and natural language, is considered to explore its ability to reconstruct structured high-level code from low-level inputs.
\end{itemize}

We performed a stratified split between training, validation, and testing in a ratio of $90/5/5$. The dataset was partitioned using a project-level stratified split, accounting for code length distribution. This rigorous separation avoids cross-project contamination and ensures that performance metrics are not skewed by variations in function size. Based on the previously discussed size discrepancy, we constrained the decompilation output to half the size of the input assembly representation. The maximum input sequence length (assembly) was fixed at $512$ tokens, while the maximum output sequence length (C code) was set to $256$ tokens, resulting in a dataset reduction of $\sim50\%$. Samples exceeding the predefined token limits were discarded to prevent code truncation and to limit computational overhead. The selected thresholds ensure a trade-off between representational expressiveness and computational feasibility.
To standardize training conditions across experiments, all neural models were optimized using common hyper-parameters.
Model-specific hyper-parameters, such as the number of training epochs, batch size, beam size, and evaluation frequency, were adapted according to architectural complexity, memory constraints, and expected performance. Detailed training configurations for each architecture (x86 and ARM) are reported in the Appendix \ref{app:setup}.

\subsection{Vulnerability Detection Setup}
For the vulnerability detection phase, we adopted several models pretrained on large-scale code corpora. Specifically, we evaluated Qwen2.5-Coder 3B \citep{hui2024qwen25codertechnicalreport}, DeepSeek-Coder-Instruct 1.3B \citep{guo2024deepseekcoderlargelanguagemodel}, CodeT5 \citep{wang2021codet5}, CodeBERT \citep{feng2020codebert}, CodeGPT \citep{radford2019gpt2}, CodeParrot \citep{face2023codeparrot}, GraphCodeBERT Base \citep{guo2021graphcodebertpretrainingcoderepresentations}, and PLBART \citep{plbart}. GraphCodeBERT extends the capabilities of CodeBERT by incorporating the inherent structure of source code. Unlike traditional models that treat code as a simple sequence of tokens, GraphCodeBERT leverages data flow graphs. 

To ensure architectural consistency for the classification task, encoder-only models (e.g., CodeBERT, GraphCodeBERT) were directly fine-tuned with a classification head. For encoder–decoder models (e.g., CodeT5 and PLBART), we followed the approach established by \cite{wang2021codet5}, retaining only the encoder component to extract sequence representations. For decoder-only models (e.g., CodeGPT and CodeParrot, based on the GPT architecture), we employed the standard sequence classification variant established by \cite{radford2019gpt2}, where prediction is performed using the hidden state of the last non-padding token.

All models are fine-tuned for source code vulnerability classification. Training data consists of a blend of VulDec and unused samples from DiverseVul \citep{chen2023diversevul} external to VulDec and the end-to-end evaluation. This integration aims to enrich data diversity while avoiding redundant instances.

To ensure consistency with the decompilation pipeline and enable a fair comparison between the source and decompiled code, input sequences were constrained to a maximum length of 256 tokens. Functions with fewer than 20 tokens or exceeding this limit were discarded. As a result, the dataset was reduced by $\sim25\%$. This filtering step significantly reduced the number of vulnerable instances and further exacerbated the original class imbalance ($3\%$ vulnerability rate).

To address the severe class imbalance of the dataset, we introduced a validation phase aimed at identifying the \textit{optimal imbalance mitigation strategy}. For each model, we benchmarked different mitigation techniques on the validation set, which was intentionally kept imbalanced ($3\%$ vulnerability prevalence) to mirror a realistic distribution. We evaluated the impact of standard training (\textit{Raw}), cost-sensitive learning via \textit{Class Weighting} (\textit{CW}), and various configurations of \textit{Under-Sampling} (\textit{US}) aimed at altering the training distribution.
The configuration that yielded the best trade-off between minority-class precision and recall on the validation set was then selected as the definitive training pipeline. The dataset is partitioned into training, validation, and test sets using a stratified $90/5/5$ split to preserve class balance across all subsets. Further details on hyper-parameter tuning and training splits are provided in the Appendix \ref{app:setup}.
\section{Experimental Evaluation}
\label{sec:experimental_eval}

In this section, we evaluate the performance of our approach by analyzing its capabilities in the decompilation and vulnerability detection tasks.

\subsection{Decompilation Task}

The performance of LLMs was assessed using a set of complementary metrics, summarized in Table \ref{tab:neural_decompilation_eval}. We employed Edit Similarity (ES), BLEU-4, and ROUGE-L to quantify translation quality. The Edit Distance (ED), adopted by \citet{hosseini2022beyond} and \citet{katz2018using}, measures structural fidelity by counting the minimal edits (insertions, deletions, and substitutions) needed to transform the decompiled output into the reference code. ES is a normalized version of ED in the range [0,1]. Given two token sequences A and B, its formula is
\begin{equation}
    ES(A, B) = 1 - \frac{ED(A, B)}{\max(|A|,|B|)},
\end{equation}
where $|A|$ and $|B|$ are the lengths of the two sequences. The higher its value, the better the translation will be. BLEU-4 evaluates n-gram precision, and ROUGE-L captures structural and semantic overlaps. However, metrics such as BLEU, originally developed for natural text translation, may not be particularly appropriate for evaluating translation between code sequences \citep{liguori2023evaluates, tran2019does}. Although originally designed for natural language, these metrics provide useful insights into code translation, with ES being particularly suited for capturing fine-grained differences.

\begin{table*}[t]
\centering
\caption{Automatic evaluation of the neural decompilation task. Results are reported for x86 and ARM architectures}
\label{tab:neural_decompilation_eval}

\resizebox{0.95\textwidth}{!}{
\begin{tabular}{lccccc}
\toprule

\textbf{Model}
& \textbf{\# parameters}
& \textbf{ISA}
& \textbf{Edit Similarity}
& \textbf{BLEU-4}
& \textbf{ROUGE-L}
\\

\midrule

\textbf{Qwen2.5-Coder}
& 3.0B
& x86
& \textbf{55.83}
& 40.50
& \textbf{56.45}
\\
& 
& ARM
& \textbf{56.60}
& 43.42
& \textbf{55.96}
\\

\midrule

CodeT5+
& 220M
& x86
& 55.55
& \textbf{42.45}
& 37.34
\\
&
& ARM
& 56.12
& \textbf{44.41}
& 38.10
\\

\midrule

DeepSeek-Coder
& 1.3B
& x86
& 49.18
& 29.76
& 45.97
\\
&
& ARM
& 47.17
& 27.51
& 44.21
\\

\midrule

CodeBERT
& 125M
& x86
& 47.39
& 26.27
& 35.16
\\
&
& ARM
& 43.56
& 20.46
& 34.27
\\

\midrule

PLBART
& 140M
& x86
& 34.29
& 14.34
& 25.19
\\
&
& ARM
& 27.19
& 7.41
& 19.66
\\

\midrule

CodeGPT
& 124M
& x86
& 23.96
& 2.60
& 20.78
\\
&
& ARM
& 22.04
& 0.16
& 19.35
\\

\midrule

CodeParrot
& 110M
& x86
& 27.14
& 3.95
& 24.11
\\
&
& ARM
& 24.67
& 2.94
& 22.14
\\

\bottomrule
\end{tabular}
\label{tab:Neural decompilation evaluation}
}
\end{table*}

Across both x86 and ARM datasets, Qwen2.5-Coder \citep{hui2024qwen25codertechnicalreport} achieves the strongest overall performance, reaching an Edit Similarity of 55.83\% on x86 and 56.60\% on ARM, while dominating the token-level context coverage with a ROUGE-L score of 56.45\% (x86) and 55.96\% (ARM). Interestingly, \textit{DeepSeek-Coder} \citep{guo2024deepseekcoderlargelanguagemodel}, despite having fewer than half the parameters of Qwen, demonstrates highly competitive token-generation capabilities, scoring a remarkably high ROUGE-L of 45.97\% on x86 and 44.21\% on ARM. This underscores the substantial impact of scaling and advanced code-specific pre-training recipes.

A notable finding is the performance of CodeT5+ \citep{wang2023codet5}. Despite being an order of magnitude smaller than Qwen2.5-Coder, it achieves the highest overall BLEU-4 scores (42.45\% on x86 and 44.41\% on ARM) and highly competitive Edit Similarity values. This exceptional efficiency is highly attributable to its native encoder-decoder architecture and span-denoising pre-training objective. PLBART \citep{plbart} ranks intermediate: its multi-task denoising improves robustness to corrupted inputs, yet it falls short of CodeT5 in sequence-level accuracy.

Small, pure autoregressive models, i.e., CodeGPT \citep{radford2019gpt2} and CodeParrot \citep{face2023codeparrot}, fail drastically on this task. This failure underscores that generation-only architectures, when lacking sufficient capacity or bidirectional context during encoding, cannot effectively map low-level structural assembly sequences to high-level source code. Conversely, encoder-only models adapted to a seq-to-seq pipeline, i.e., CodeBERT \citep{feng2020codebert}, maintain much higher structural robustness, retaining an Edit Similarity above 43\%, proving that robust contextual representation of assembly instructions is a critical prerequisite for meaningful translation.

The results further reveal a non-uniform influence of the target ISA on performance. For CodeT5 and Qwen, ARM binaries slightly outperform x86 across most metrics, suggesting a more regular or learnable instruction-to-source mapping. By contrast, DeepSeek, CodeBERT, and PLBART exhibit a more pronounced degradation on ARM, indicating greater sensitivity to architectural variation. Overall, however, no consistent trend clearly favors one ISA over the other, although the ARM architecture instruction set displays marginally lower syntactic complexity than x86, consistent with its RISC principles of fixed-length instructions, limited addressing modes, and a highly regular operand structure.

Table \ref{tab:humaneval} reports the performance on the HumanEval benchmark \citep{humaneval} under \texttt{-O2} optimization, comparing our VulDec Decompiler against rule-based decompilation, Ghidra \citep{Ghidra}, and the specialized \textit{LLM4Decompile} family \citep{Tan_2024}.

\begin{table*}[htbp]
\caption{Human-Eval BenchMark (-opt 02) on VulDec x86 Decompiler}
\begin{center}
\resizebox{0.95\textwidth}{!}{%
\begin{tabular}{llcccccc}
\toprule
\textbf{Framework} & \textbf{Model} & \textbf{ES} & \textbf{Re-executability}\\
\midrule
Ghidra & Base & 6.19 & 15.24\\
\midrule
\multirow{ 3}{*}{\cite{Tan_2024}}  
& LLM4Decompile-End-6.7B & 12.93 &  36.71  \\
& Ghidra + LLM4Decompile-Ref-6.7B & 13.42 &  47.56  \\
& Ghidra + LLM4Decompile-Ref-22B & 13.58 &  \textbf{59.76}  \\
& Ghidra + LLM4Decompile-Ref-33B & 13.63 &   43.90  \\
\midrule
VulDec Decompiler & Qwen2.5-Coder-3B & \textbf{32.82} & 34.34  \\
\bottomrule
\end{tabular} 
}%
\label{tab:humaneval}
\end{center}
\end{table*}

In terms of structural alignment, our framework yields a substantial improvement, reaching an edit similarity of 32.82\%, which noticeably exceeds the scores of the other evaluated models. While hybrid multi-stage pipelines like \textit{Ghidra + LLM4Decompile-Ref-22B} achieve higher re-executability (59.76\%), VulDec achieves a competitive 34.34\%, performing closely to the standalone \textit{LLM4Decompile-End-6.7B} (36.71\%) despite utilizing a significantly smaller model (3B parameters) and operating without external decompilation outputs. As expected, the baseline Ghidra exhibits a low Edit Similarity of 6.19\%. This occurs because rule-based decompilers focus on functional reconstruction rather than matching human coding styles, often generating nested structures, unpromoted types, and generic variable names that fail to compile directly or match the reference source text. 

\subsection{Vulnerability Detection Task}

Table \ref{tab:strategy} summarizes the validation results for the encoder-only \textit{CodeBERT} model to identify the optimal class-imbalance mitigation strategy. To ensure a valid assessment, all configurations are evaluated on a validation set that preserves the operational distribution ($3\%$ vulnerability prevalence). We benchmarked several \textit{under-sampling} (US) configurations, systematically varying the majority-to-minority class ratios, namely 50:50, 70:30, and 80:20. In tandem, we evaluated a cost-sensitive \textit{class weighting} (CW) scheme designed to penalize minority-class misclassifications. Specifically, the weight assigned to each class was defined as inversely proportional to its training prevalence; for instance, under the 80:20 US configuration, the resulting class weights are set to 1:4 for the majority and minority classes, respectively.
\begin{table*}[htbp]
\centering
\caption{Class-imbalance mitigation strategies using CodeBERT on validation set ($3\%$ vulnerability prevalence). CW: Class Weighting; US: Under-Sampling}
\label{tab:strategy}

\resizebox{0.9\textwidth}{!}{
\begin{tabular}{lcccc}
\toprule

\textbf{Strategy}
& \textbf{Vul-Prec}
& \textbf{Vul-Recall}
& \textbf{Vul-F1}
& \textbf{F1-score (macro avg)}
\\

\midrule

Raw
& 0.463
& 0.097
& 0.160
& 0.571
\\

Raw + CW
& \textbf{0.532}
& 0.115
& 0.189
& 0.586
\\

\midrule

US 50-50
& 0.070
& 0.707
& 0.128
& 0.466
\\

US 70-30
& 0.127
& 0.427
& 0.196
& 0.567
\\

US 80-20
& 0.167
& 0.336
& 0.223
& 0.591
\\

\midrule

\textbf{US 80-20 + CW}
& 0.171
& \textbf{0.415}
& \textbf{0.243}
& \textbf{0.599}
\\

\bottomrule
\end{tabular}
}
\end{table*}

The empirical results highlight a sharp trade-off between precision and recall. Standard training configurations (\textit{Raw} and \textit{Raw + CW}) under-represent the minority class, leading to negligible recall. Conversely, a perfectly balanced training distribution, US 50-50, over-compensates, causing an unacceptable collapse in precision ($0.070$) and a severe degradation of the macro F1-score. The data indicate that shifting towards moderate under-sampling ratios stabilizes the model. For \textit{CodeBERT}, the best performance across all integrated metrics is achieved by the hybrid \textbf{US 80-20 + CW} strategy, which yields the top \textit{Vul-Recall} ($0.415$) and \textit{Macro F1-score} ($0.599$). We extended this validation setup to one model per architecture family in Appendix \ref{app:strategy}. The US 80-20 + CW configuration consistently proved to be the optimal choice for the majority of architectures. However, for \textit{Qwen2.5-Coder} and \textit{DeepSeek-Coder} architectures, the balanced \textbf{US 50-50} ratio yielded the highest overall performance, suggesting a higher tolerance for class balance adjustments. Consequently, each model was trained using its respective optimal configuration in the subsequent evaluation phases.

\begin{table*}[htbp]
\centering
\caption{Vulnerability prediction performance on an imbalanced in-distribution test set ($3\%$ vulnerable samples). 
Reported metrics refer to the vulnerable class and overall macro performance}
\label{tab:vuln_results}

\resizebox{0.9\linewidth}{!}{
\begin{tabular}{lccccccc}
\toprule

\multicolumn{8}{c}{\textbf{In-distribution Test Set}}
\\

\midrule
\textbf{Model}
& \textbf{Vul-P}
& \textbf{Vul-R}
& \textbf{Vul-F1}
& \textbf{Acc}
& \textbf{Prec}
& \textbf{Recall}
& \textbf{F1}
\\

\midrule

DiverseVul$^{*}$
& --
& --
& --
& 0.923
& 0.518
& 0.433
& 0.472
\\

FlawFinder
& 0.055            
& 0.130
& 0.077
& 0.898 
& 0.512        
& 0.527
& 0.511
\\

\midrule

% soglia 0.75, 50-50

CodeT5 (base)
& 0.150
& 0.309
& 0.202
& 0.920
& 0.563
& 0.625
& 0.580
\\

CodeParrot
& 0.164
& 0.276
& 0.206
& 0.930
& 0.569
& 0.614
& 0.584
\\

DeepSeek-Coder
& 0.180    
& 0.260
& 0.213
& \textbf{0.937}
& 0.577
& 0.610
& 0.590
\\

Qwen2.5-Coder
& 0.180
& 0.301
& 0.226
& 0.932
& 0.579
& 0.627
& 0.596
\\
% soglia 0.75, 50-50

CodeGPT
& 0.160
& 0.417
& 0.231
& 0.909
& 0.569
& 0.671
& 0.591
\\

GraphCodeBERT
& 0.168
& 0.423
& 0.240
& 0.912
& 0.574
& 0.676
& 0.597
\\

CodeBERT
& 0.176
& 0.398
& 0.244
& 0.919
& 0.577
& 0.667
& 0.601
\\

\midrule

\textbf{PLBART}
& \textbf{0.180}
& \textbf{0.444}
& \textbf{0.257}
& 0.916
& \textbf{0.580}
& \textbf{0.688}
& \textbf{0.606}
\\

\bottomrule
\multicolumn{8}{l}{\small $^{*}$ Trained on an extended dataset incorporating additional external sources}
\end{tabular}
}
\end{table*}

Table \ref{tab:vuln_results} reports model performance on \textit{project-level in-distribution} (ID) data, i.e., code from projects already seen during training for the vulnerability detection task. To ensure a realistic operational evaluation, all architectures are tested on an imbalanced test set with a $3\%$ vulnerability prevalence. Models are evaluated using standard metrics (Accuracy, Precision, Recall, Macro F1) alongside minority-class specific metrics (\textit{Vul-Precision}, \textit{Vul-Recall}, and \textit{Vul-F1}) to assess their behavior on the heavily imbalanced minority.
The results show that overall \textit{Accuracy} is an uninformative metric in this highly imbalanced scenario; for instance, the baseline DiverseVul \citep{chen2023diversevul} classifier, despite having been trained on an extended dataset incorporating additional external data sources, yields a high accuracy ($0.923$) but completely fails to effectively capture the minority class, resulting in an exceptionally low macro F1-score ($0.472$). The rule-based static analyzer, \textit{FlawFinder} \citep{flawfinder}, exhibits poor capabilities in identifying vulnerable samples, yielding a critical \textit{Vul-F1} of $0.077$.

Overall, the absolute performance metrics across all evaluated neural architectures remain relatively modest, with the highest vulnerability-specific F1-score (\textit{Vul-F1}) reaching only $0.257$. This concern highlights the intrinsic complexity and persistent challenges of software vulnerability detection under severe real-world class imbalance, an open problem widely documented in the literature \citep{secvuleval, arewethereyet,lprotector}.

Among the evaluated neural architectures, PLBART \citep{plbart} achieves the best results for the security task, maximizing both \textit{Vul-Recall} ($0.444$) and \textit{Vul-F1} ($0.257$). Encoder-based models (CodeBERT and GraphCodeBERT) track closely, maintaining a balanced trade-off with a \textit{Vul-F1} of approximately $0.24$. Conversely, decoder-only architectures, i.e., CodeParrot and CodeGPT, along with standard \textit{seq2seq} baselines CodeT5, experience a visible performance degradation on the minority class, showing lower precision and restricted recall profiles. A particularly notable trend emerges when examining larger model architectures ($>1\text{B}$ parameters). Despite their superior parameter scale, these models fail to outperform significantly smaller encoder and sequence-to-sequence networks. \textit{DeepSeek-Coder} and \textit{Qwen2.5-Coder} report \textit{Vul-F1} scores of $0.213$ and $0.226$, respectively, placing them in the lower range of learning-based models. Our observation that larger models do not strictly yield superior performance mirrors empirical findings in the literature \citep{chen2023diversevul}, where scaling model parameters similarly produced non-monotonic improvements in vulnerability detection.

\subsection{End-to-End Evaluation}

Once the core pipeline components, specifically \textit{Qwen2.5-Coder} \citep{hui2024qwen25codertechnicalreport} for decompilation and \textit{PLBART} \citep{plbart} for vulnerability detection, have been established, the proposed framework undergoes an end-to-end inference evaluation. 

The evaluation is conducted on functions extracted from projects that are disjoint from the training set, \textit{i.e., project-level out-of-distribution} (OOD), in order to ensure rigorous generalization. This test set comprises over a thousand real-world functions gathered from the following open-source projects: \texttt{PHP}, \texttt{openjpeg} and \texttt{poppler}. Furthermore, this evaluation set preserves the realistic operational distribution with a $3\%$ vulnerability prevalence, effectively exposing the framework to novel variations in coding style, API usage, and program structures. For all subsequent experiments, we focus exclusively on binaries compiled for the x86 architecture.

\begin{figure*}[!htbp]
\centering
%\centerline{\includegraphics[width=0.7\textwidth]
\centerline{\includegraphics[width=0.85\textwidth]{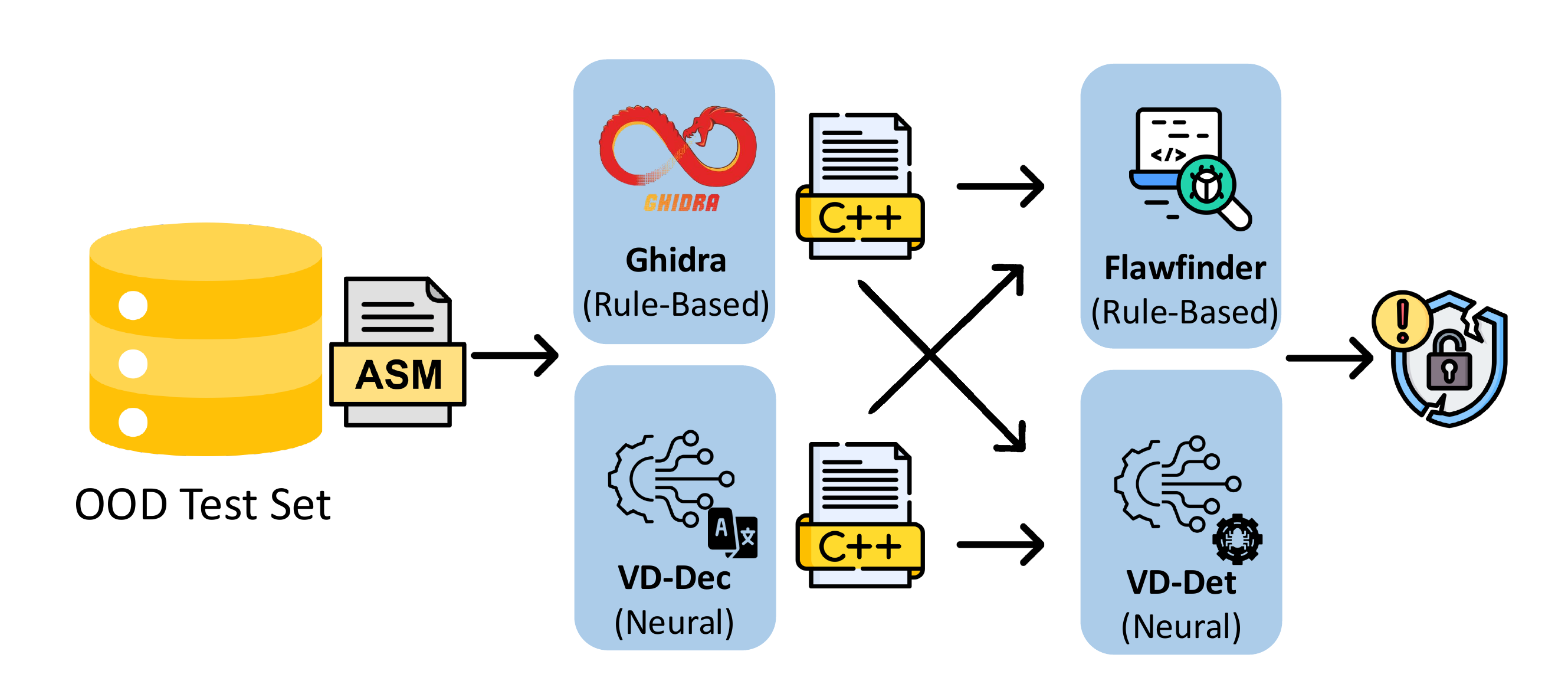}}
\caption{Architectural overview of the end-to-end evaluation via cross-combination pipelines evaluating rule-based and neural stages for decompilation and detection}
\label{fig:EndtoEnd}
\end{figure*}

We integrate two industry-standard, rule-based security and reverse-engineering tools into our evaluation. For the translation stage, we employ \textit{Ghidra} \citep{Ghidra}\footnote{Since Ghidra cannot process isolated assembly code without a complete binary structure, we built a preprocessing pipeline that leverages \textit{Keystone}  \citep{keystone_engine} to turn target function fragments back into raw executable binaries, allowing Ghidra's headless mode to decompile them.}, a widely adopted software reverse engineering framework driven by deterministic, rule-based decompilation heuristics. For the vulnerability detection stage, we adopt \textit{Flawfinder} \citep{flawfinder}, a well-established static analysis tool that scans source code for security flaws by matching rigid syntactic patterns.

To assess the joint impact of decompilation quality on downstream security tasks, we perform an end-to-end evaluation testing four distinct pipeline configurations. As illustrated in Fig. \ref{fig:EndtoEnd}, the process orchestrates a translation stage (mapping binary assembly to source code) followed by a vulnerability detection stage. Specifically, our experiments contrast the fully traditional pipeline (\textit{Ghidra $\rightarrow$ Flawfinder}) against our fully neural framework VulDec, alongside two cross-combined hybrid configurations (\textit{Ghidra $\rightarrow$} VD-Det and VD-Dec \textit{$\rightarrow$ Flawfinder}). This cross-evaluation allows us to isolate the individual contributions of both the decompilation and the detection components under realistic conditions.

\begin{table*}[htbp]
\centering
\caption{End-to-end evaluation on an imbalanced out-of-distribution test set (3\% vulnerability prevalence). The columns report cross-combinations of rule-based and neural tools for both decompilation and vulnerability detection}
\label{tab:ed45_results}

\resizebox{0.9\linewidth}{!}{
\begin{tabular}{l|cc|cccc}
\toprule

\multicolumn{7}{c}{\textbf{Out-of-distribution Test Set}}
\\

\midrule

\textbf{Pipeline} 
& \textbf{ES}
& \textbf{BLEU-4}
& \textbf{Vul-P}
& \textbf{Vul-R}
& \textbf{Vul-F1}
%& \textbf{Acc}
& \textbf{F1}
\\

\midrule

Ghidra  $\rightarrow$ Flawfinder
& 26.27
& 9.19
& 0.00
& 0.00
& 0.00
%& 0.94
& 0.49
\\

Ghidra $\rightarrow$ VD-Det
& 26.27
& 9.19
& 0.03 %0.07 
& 0.04 %0.09 
& 0.03 %0.08 
%& 0.88
& 0.50 
\\

VD-Dec $\rightarrow$ Flawfinder
& \textbf{36.83} %35.35
& \textbf{19.36} %18.37
& 0.04 %0.05
& 0.07 %0.07
& 0.05 %0.05
%& 0.90
& 0.51
%& 0.07 %0.11
%& 0.06 %0.09
%& 0.06 %0.10
%& 0.92 %0.91
%& 0.51 %0.53
\\

\midrule

VD-Dec $\rightarrow$ VD-Det
& \textbf{36.83} %36.56 35.35
& \textbf{19.36} %18.87 18.37
& \textbf{0.14} %0.13 0.12
& \textbf{0.28} %0.36 0.24
& \textbf{0.19} %0.19 0.16
%& 0.89 %0.85
& \textbf{0.57} %0.54
\\

\bottomrule
\end{tabular}
}
\end{table*}

Table \ref{tab:ed45_results} reports the performance across both translation metrics (Edit Similarity, BLEU-4) and classification metrics.
The end-to-end evaluation yields several critical insights outlined below.

\vspace{2pt}
\noindent
\textbf{End-to-End Pipeline Analysis and Cross-Evaluation.} The traditional rule-based pipeline, i.e., \textit{Ghidra $\rightarrow$ Flawfinder}, fails entirely in this realistic setup and collapses on the minority class, reporting a Vul-F1 of $0.00$. This highlights the inability of rigid syntactic pattern-matchers to generalize over the structural variations generated during standard decompilation. This architectural limitation is slightly resolved by our neural decompiler, which establishes a superior baseline over Ghidra. It shows a strong regularizing effect on downstream tools. Replacing Ghidra with VD-Dec allows even a traditional analyzer, i.e., Flawfinder, to capture vulnerabilities ($\text{Vul-F1} = 0.05$). The fully neural pipeline, \textit{VD-Dec $\rightarrow$ VD-Det}, achieves the best performance, maximizing Vul-Recall ($0.28$) and tracking the highest global F1-score.

\vspace{2pt}
\noindent
\textbf{The Generalization Gap: ID vs. OOD Robustness.}
A comparative analysis between the evaluation on the in-distribution test set and the out-of-distribution set reveals an \textit{evident performance degradation}, particularly affecting the minority-class recall, which drops from $0.44$ to $0.28$ for the optimal configuration. 
%This performance gap is driven by several compounding technical dimensions. 

This behavior exemplifies a classic \textit{out-of-distribution generalization problem}, a well-documented limitation in the current state-of-the-art. Recent benchmarks, such as \cite{ood}, prove that high in-distribution performance is often a misleading metric driven by spurious correlations, showing severe accuracy drops, e.g., from $0.70$ to $0.49$, when models are exposed to real-world OOD test sets. Similarly, \cite{arewethereyet} highlights that evaluating deep learning models on real-world test sets with different distributions triggers a disastrous average performance drop of $\sim -73\%$. The DiverseVul classifier \citep{chen2023diversevul} suffers a massive architectural collapse under a comparable domain shift, with its reported Macro F1-score dropping from $0.472$ on the ID test set to a near-baseline $0.110$ on the OOD test set. To further underline this vulnerability, frameworks like \textit{SLODA} \citep{attackood} even exploit OOD features to intentionally mislead code models via adversarial attacks. More broadly, while an ID test set evaluates models on familiar software patterns encountered during training, an OOD set forces them to generalize over entirely novel coding conventions and library abstractions. This fundamental limitation leaves significant room for future optimization in the community.

Furthermore, the structural nature of the input code changes drastically between the two experimental setups: the ID evaluation operates directly on \textit{source code}, whereas this OOD end-to-end framework forces the detector to ingest \textit{noisy, decompiled code}. Intuitively, the model would be expected to struggle with the syntactic artifacts and missing symbols introduced during translation. However, as shown in the following section, empirical results reveal a counterintuitive trend, proving that this noise is not the primary cause of the performance gap.

%The challenge is heavily exacerbated by the realistic, highly \textit{skewed class distribution}. Under such strict conditions, even minor semantic ambiguities in the decompiled code lead the model to safely default to the dominant majority class.

\subsection{Impact of decompilation on vulnerability prediction}

\begin{figure}[htbp]
\centering
\includegraphics[width=1\linewidth]{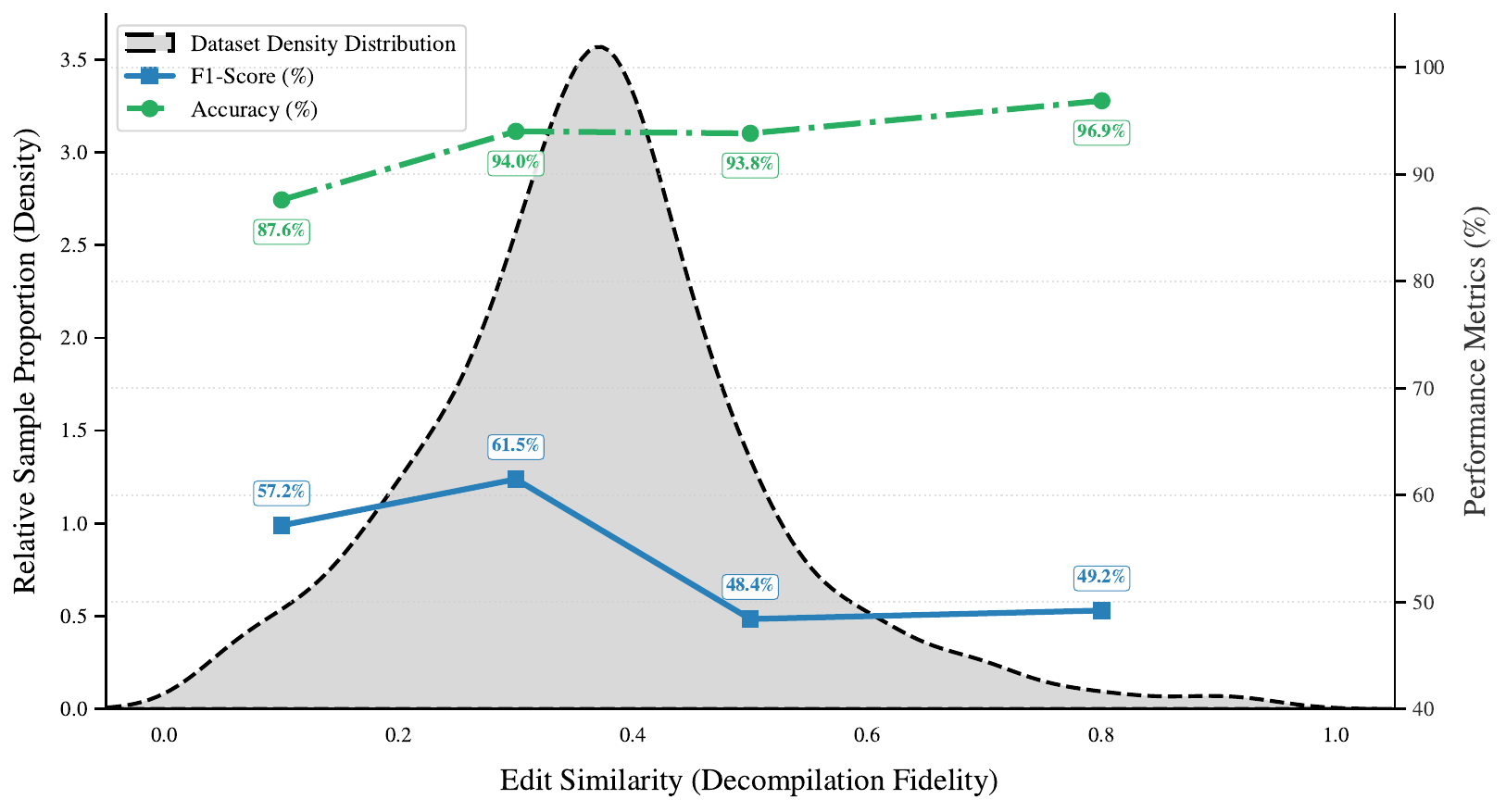}
\caption{Correlation analysis between Edit Similarity (Decompilation Fidelity) and downstream classification performance (Accuracy and F1-Score). The density curve (left) highlights the sample distribution across the test set, while the trend lines (right) illustrates the performance metrics values as structural translation quality increases}
\label{fig:correlation_analysis}
\end{figure}

To further investigate the relationship between translation quality and security classification, Fig. \ref{fig:correlation_analysis} correlates the \textit{Decompilation Fidelity (Edit Similarity)} with the downstream \textit{Classification Accuracy} and \textit{Macro F1-Score}.

 %We observe a clear, monotonic upward trend: the classification accuracy increases alongside the growth of the Edit Similarity (ES). While global accuracy can be an optimistic metric under severe class imbalance, its evaluation across localized fidelity intervals provides a reliable indicator of the model's stability. Crucially, the density distribution highlights that the majority of samples cluster around a fidelity of $0.36$, where the pipeline stabilizes at an accuracy of $85\%$ matching the aggregate results in Table \ref{tab:ed45_results}. The continuous growth in the low-density tail ($0.6 \le ES \le 1.0$) empirically validates our core hypothesis: shielding the detector from structural and syntactic translation noise directly enhances its decision-making consistency.

As illustrated in Figure~\ref{fig:correlation_analysis}, the dataset density distribution reveals that the vast majority of samples cluster around an Edit Similarity (ES) value of approximately $0.36$. Visually, while downstream Accuracy follows a clear monotonic upward trend with increasing ES, the vulnerability detection F1-score exhibits a non-monotonic pattern, peaking at $61.5\%$ near an ES of $0.3$ before stabilizing around $49\%$.
 
\vspace{2pt}
\noindent
\textbf{Correlation Analysis.} To rigorously validate these empirical observations, we performed a statistical correlation analysis examining the relationships between Edit Similarity, Accuracy, F1-score, and code scale using Spearman’s rank ($\rho$) and Point-Biserial correlation tests at a $95\%$ confidence level ($\alpha = 0.05$). The results reveal a \textit{clear decoupling between syntactic fidelity and security semantics}. Specifically, while Accuracy significantly correlates with structural quality, the F1-score shows no statistically meaningful dependency on Edit Similarity. This finding means that generating syntactically or structurally similar code does not inherently help the downstream model in capturing subtle security vulnerabilities. Furthermore, code scale acts as a penalty on generation performance: a larger function length (higher token count) significantly degrades both Edit Similarity and Accuracy, yet displays no statistically significant impact on the vulnerability detection F1-score.

\vspace{2pt}
\noindent
\textbf{Control Experiment.} Finally, we conducted a control experiment by evaluating the detector directly on the \textit{source code} of the OOD test set in order to rigorously isolate whether the observed OOD performance degradation stems from decompilation noise. Strikingly, the model yielded nearly \textit{identical results}, with a negligible variance of $\pm 0.02$ (maintaining a vulnerable-class recall of $0.28$ and precision of $0.14$). The control experiment alignment suggests that the decompilation process is not the primary driver of the performance drop, as the model faces the same difficulties even when analyzing clean source code. This result indicates that the detector has reached a generalization plateau; rather than capturing deep, universal security semantics, the LLM appears to rely on stylistic and syntactic details specific to the training distribution, leading to the same modest performance.

\subsection{Effectiveness of VulDec Across Vulnerability Categories}

The behavior of the vulnerability detection model was analyzed across vulnerability categories to assess potential category-level imbalances. The binary detection evaluation (safe vs. vulnerable) was extended to a multiclass setting using CWE annotations, allowing for the identification of both vulnerable code and specific weakness types. The VulDec dataset comprises 148 distinct CWE codes, some of which are missing. To reduce computational and interpretative complexity and to group similar vulnerability types, functions were consolidated into several macro-categories shown in Table \ref{tab:vuln_group_performance_distribution}. The assignment, reported in Appendix \ref{app:map}, follows a custom taxonomy inspired by MITRE categories \citep{Mitre} and is designed to ensure that each function is annotated with its binary status, CWE identifier (when available), and the corresponding vulnerability group. Metrics are computed based on PLBART's evaluation of the in-distribution test set.

\begin{table}[htbp]
\caption{Performance of vulnerability prediction across vulnerability groups reported in Appendix \ref{app:map} and ID test set distribution}
\centering
%\resizebox{\columnwidth}{!}{%
\begin{tabular}{l|c|cccc}
\toprule
\textbf{Group} &
\textbf{Distr \%} &
\textbf{Prec} &
\textbf{Rec} &
\textbf{Acc} &
\textbf{F1} \\
\midrule
 Uncategorized                       &  48.47 & 0.58 & 0.68 & 0.91 & 0.60 \\
 Memory Management and Concurrency   &  32.30 & 0.58 & 0.72 & 0.92 & 0.61 \\
 Validation and Authorization Issues &   6.66 & 0.60 & 0.72 & 0.92 & 0.63 \\
 Taint Propagation Issues            &   7.89 & 0.59 & 0.67 & 0.94 & 0.62 \\
 Data Protection Issues              &   4.19 & 0.53 & 0.57 & 0.91 & 0.54 \\
 Insecure Configuration Issues       &   0.48 & 0.49 & 0.42 & 0.83 & 0.45 \\
\bottomrule
\end{tabular}
%}
\label{tab:vuln_group_performance_distribution}
\end{table}
\begin{comment}
    Memory Management and Concurrency
 & 37.44 & 0.75 & 0.68 & 0.74 & 0.71 \\
Data Protection Issues
 & 22.60 & 0.72 & 0.81 & 0.74 & 0.76 \\
Taint Propagation Issues
 & 16.78 & 0.70 & 0.68 & 0.69 & 0.69 \\
Uncategorized
 & 15.07 & 0.73 & 0.66 & 0.73 & 0.69 \\
Validation and Authorization Issues
 & 5.25 & 0.58 & 0.61 & 0.71 & 0.59 \\
Insecure Configuration Issue
 & 2.85 & 0.64 & 0.54 & 0.67 & 0.59 \\
\end{comment}

The distribution of functions across macro-categories reveals a marked imbalance, with \textit{Uncategorized} samples and \textit{Memory Management and Concurrency} dominating the dataset ($48.47\%$ and $32.30\%$ prevalence, respectively). This layout aligns with traditional C/C++ repository distributions, where memory-related flaws represent the historical baseline. Conversely, categories like \textit{Data Protection Issues} ($4.19\%$) and \textit{Insecure Configuration Issues} ($0.48\%$) are severely underrepresented. 

Performance, detailed in Table \ref{tab:vuln_group_performance_distribution}, varies across these families when evaluated through macro-averaged metrics. \textit{Validation and Authorization Issues} achieved the highest localized detection capability, reaching a Macro Recall of $0.72$ and a Macro F1-score of $0.63$, closely followed by \textit{Memory Management and Concurrency} ($0.72$ Recall, $0.61$ F1-score). This suggests that the structural and contextual cues tied to buffer tracking and access control loops provide robust, recognizable patterns for the downstream detector.

In contrast, \textit{Data Protection Issues} and \textit{Insecure Configuration Issues} present the weakest macro performance. For the latter, the extreme scarcity of data ($0.48\%$ distribution) severely constrains the model's predictive alignment, dragging the Macro F1-score to $0.45$ and the Macro Recall to $0.42$. 

While \textit{Accuracy} remains consistently high ($0.83$ to $0.94$), it is mathematically padded by the massive volume of non-vulnerable samples correctly identified by the model. The macro-averaged F1-score, settling around $0.60$ for the main categories, provides a far more reliable proxy of the actual trade-off within each macro-group.

\subsection{Assembly-Based Vulnerability Detection}

To evaluate the impact of bypassing the decompilation stage, we trained a CodeBERT-based classifier to perform vulnerability detection directly on native x86 assembly using the VulDec dataset. This configuration, shown in Figure \ref{fig:asm}, establishes a direct assembly-to-vulnerability pipeline, effectively removing the intermediate abstraction layer. For training and subsequent evaluation, we adopted the same careful sample pre-processing procedure performed in Section \ref{sec:experimental_setup}.
\begin{table}[htbp]
\centering

\begin{minipage}{0.45\textwidth}
    \centering
    \caption{Assembly-based vulnerability detection on VulDec dataset}
    \label{tab:ass_results}
    \vspace{0.2cm}
    \small
    \begin{tabular}{c | c c c}
    \toprule
    \textbf{Class} & \textbf{Precision} & \textbf{Recall} & \textbf{F1-Score} \\
    \midrule
    Safe & 0.89 & 1.00 & 0.92 \\
    Vul  & 0.00 & 0.00 & 0.00 \\
    \bottomrule
    \end{tabular}
\end{minipage}
\hfill
\begin{minipage}{0.45\textwidth}
    \centering
    \includegraphics[width=\textwidth]{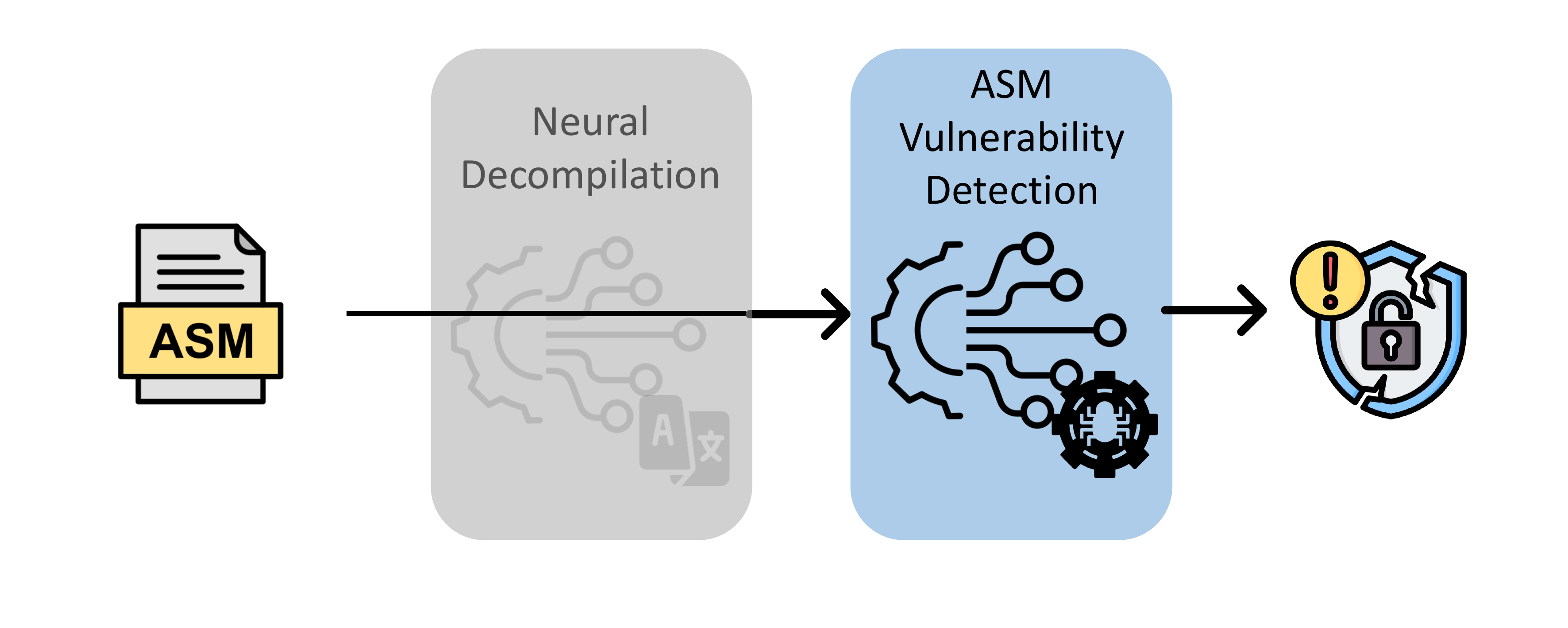}
    \captionof{figure}{Vulnerability detection on raw x86 assembly (no decompilation baseline)}
    \label{fig:asm}
\end{minipage}

\end{table}

The results, reported in Table \ref{tab:ass_results}, show strong performance in the Safe class (precision 0.89, recall 1.00, F1-score 0.92), while completely failing to identify vulnerable functions, resulting in zero precision, recall, and F1-score for the Vul class.

This behavior indicates that the model struggles to infer the higher-level semantic information required to recognize vulnerable behaviors. Assembly code lacks explicit control-flow abstractions, variable semantics, and contextual cues that are crucial for vulnerability reasoning. As a result, the model becomes overly biased toward Safe predictions, missing all vulnerable instances. These findings suggest that, despite simplifying the analysis pipeline, decompilation remains a necessary step for enabling effective vulnerability detection within VulDec.
\section{Lessons Learned}
\label{sec:lessons}

\vspace{2pt}
\noindent
(i) \textit{Applicability of neural decompilation to cybersecurity}.
The findings of this work suggest that the adoption of neural decompilation techniques for cybersecurity-related tasks constitutes a promising avenue for future research, potentially enabling new applications such as binary malware analysis. Although the decompiled C/C++ code exhibits limited syntactic fidelity, the preservation of its semantic content allows it to be effectively leveraged by learning-based approaches, which predominantly base their predictions on recurring patterns and structural features.

\vspace{2pt}
\noindent
(ii) \textit{Effectiveness of VulDec Dataset}.
The experimental results indicate that the VulDec dataset shows promise as a viable benchmark for training both decompilation and vulnerability detection models. The evaluation on OOD test set yields encouraging outcomes, indicating the effectiveness of the proposed methodology. Nevertheless, the performance degradation observed during the training phase highlights the limitations imposed by the adopted computational resources and model capacity, suggesting that the use of more advanced architectures and greater computational power could further enhance the framework’s effectiveness. Furthermore, the joint availability of low-level representations, high-level source code, and vulnerability annotations provided by the dataset opens up several promising research directions that lie beyond the scope of the contributions presented in this work.

\vspace{2pt}
\noindent
(iii) \textit{VulDec Framework Deployment Context}. Our empirical results indicate that while the framework achieves discrete performance, its practical utility is best realized as a \textit{first-pass screening tool} to support developers during preliminary code reviews, in line with observations by \cite{usercentered} and \cite{boundary}. Rather than acting as an autonomous triage system, the tool is intended to highlight potential high-risk regions to guide manual inspection, thereby reducing the search space for human experts. This lightweight decision-support role aligns well with the requirements of assistant tools in software maintenance and reverse engineering.
\noindent
\section{Related Work}
\label{sec:related}
This work builds upon our previous exploratory study by \cite{grasso}, in which we introduced a preliminary pipeline for binary vulnerability detection based on neural decompilation. While this study demonstrated the feasibility of the approach, it relied on synthetically generated data, specifically the Juliet dataset \citep{Juliet}, thus limiting its applicability to real-world software. In contrast, the present work addresses this limitation by introducing VulDec, an ad-hoc dataset constructed from real-world projects, and by extending the proposed pipeline to support multiple architectures, including both x86 and ARM, whereas the previous study was restricted to x86 binaries.

Neural decompilation is an innovative research trend in binary analysis. 
Recent work has explored the use of learning-based approaches for binary decompilation. \citet{hosseini2022beyond} introduced \textit{Beyond The C}, a neural decompiler based on the fairseq transformer. This solution is retargetable for several high-level programming languages, such as C, OCaml, Go, and Fortran. The results of their study are promising for all languages, scoring the best result in terms of ED ($54\%$) on the decompilation of C/C++ code (training on $2,000,000$ samples). \citet{katz2018using} proposed an approach for neural decompilation using Recurrent Neural Networks. The solution was trained and evaluated on snippets of binary machine code compiled from C source code. This solution achieved an ED equal to $30\%$, training on about $700,000$ samples. \citet{Tan_2024} introduced a family of large-scale LLMs called \textit{LLM4Decompile}, trained to translate binary code into high-level source code, achieving substantial improvements in readability and re-executability over traditional tools such as Ghidra \citep{Ghidra} and general-purpose models. 

While the previously mentioned works \citep{hosseini2022beyond,katz2018using,Tan_2024} focus on improving the quality of decompiled code as an end goal, our work differs by leveraging neural decompilation as an intermediate stage for downstream cybersecurity tasks, specifically black-box binary vulnerability detection, and by introducing a curated dataset that jointly aligns assembly, source-level semantics, and fine-grained vulnerability annotations. Existing approaches to binary vulnerability detection vary widely in terms of data sources, intermediate representations, learning paradigms, and evaluation methodologies, which makes direct experimental comparison across studies inherently challenging. \citet{zheng2019recurrent} devised an approach to identify vulnerabilities in binary files using deep learning models, particularly Recurrent Neural Networks (RNN). To this end, this work leverages the Juliet C/C++ dataset \citep{Juliet}. Then, the binary code is lifted to LLVM \citep{llvm} intermediate representation (IR), constituting the input for the RNNs. This solution can identify 4 CWEs with 90\% accuracy. \citet{schaad2023deep} extended the previous approach \citep{zheng2019recurrent}, leveraging several DL models, i.e., Recurrent Neural Networks (RNN), Long-Short-Term Memory (LSTM), and Gated Recurrent Unit (GRU). Their solution can identify if a binary file is vulnerable with 88\% accuracy, and classify 23 CWEs out of the 118 described in the Juliet dataset with 77\% accuracy. 
\citet{pewny2015cross} designed a cross-architecture solution to identify bug signatures in binary executables: the binaries are lifted to IR to unify the different assembly languages and then converted to control flow graphs (CFG) to compare basic blocks to the known signatures. This solution can identify 5 technology-specific vulnerabilities across three architectures (ARM, MIPS, x86), including the well-known \textit{Heartbleed} vulnerability for OpenSSL \citep{HeartbleedCVE, Heartbleed}. 
\emph{DiscovRE} \citep{eschweiler2016discovre} is a prototype for identifying binary code bugs. It can identify two vulnerabilities, i.e., \textit{Heartbleed} and \textit{POODLE} for OpenSSL and SSLv3 respectively, and supports four architectures, namely x86, x64, ARM, and MIPS. DiscovRE measures structural similarities between the target code and the vulnerable prototype through custom numeric features and metrics to identify the vulnerabilities. \citet{dahl2020stack} developed a solution to identify stack-based buffer overflow vulnerabilities in binary files, along with a custom dataset for this task. First, the binary code is disassembled and then compressed to discard redundant information. Again, the vulnerability detection model is an LSTM, which achieves 90\% accuracy in detecting stack-based buffer overflows.
\emph{VDISCOVER} \citep{grieco2016toward} is an ML-based tool for memory corruption detection. It comes with a custom dataset, \emph{VDISCOVERY}, built from $1,039$ programs from the Debian bug tracker. VDISCOVER can identify, with 55\% accuracy, 4 types of memory corruptions: heap memory, stack memory, corruption of an argument in a function, and corruption of a pointer to a function.

Concurrently, recent advancements in source-level vulnerability detection have shifted toward multi-agent LLM reasoning frameworks. For instance, \textit{VulTrial} \citep{widyasari2025let} uses a courtroom-inspired multi-agent paradigm with role-specific agents to debate and explain source-level bugs. Similarly, \textit{AEGIS} \citep{fang2026aegiscluesverdicts} executes multi-agent forensic verification over repository-level Code Property Graphs (CPG) to mitigate hallucinations. Crucially, the modular nature of our pipeline allows these sophisticated verification engines to be seamlessly integrated as alternative, high-fidelity downstream backends operating directly on our decompiled pseudo-C representation.

Most of the previously mentioned approaches to binary vulnerability detection rely on intermediate representations such as assembly-level tokens \citep{dahl2020stack}, LLVM IR \citep{schaad2023deep,zheng2019recurrent}, call graphs \citep{eschweiler2016discovre}, or combinations thereof \citep{pewny2015cross}. To the best of our knowledge, none of these methods exploit decompiled C/C++ pseudo-code as the primary representation. This constitutes a key distinguishing aspect of our approach, as it not only enables vulnerability prediction but also provides an interpretable approximation of the original source code from which the binary was compiled. Furthermore, many of the strong classification results reported in the literature \citep{grasso,schaad2023deep,zheng2019recurrent}, particularly in multi-class settings, are largely attributable to the use of synthetic benchmarks. Our findings \citep{grasso}, achieving up to 90\% accuracy, corroborate this observation and highlight the impact of dataset realism on reported performance.
\section{Threats To Validity}
\label{sec:threats}

\vspace{2pt}
\noindent
(i) \textit{Dataset Construction and Integrity}. A key threat to validity arises from function matching between source code and assembly representations, which relies on symbol-name equality. To evaluate alignment fidelity in the absence of independent binary verification, we conducted a manual audit on $100$ randomly sampled function pairs, confirming correct alignment in $95\%$ of cases ($95\%$ Wilson CI $[88.7\%, 98.0\%]$). This suggests reasonable empirical accuracy for symbol-based alignment, indicating that incorrect function associations introduce minimal noise. Beyond alignment integrity, since our dataset builds upon DiverseVul \citep{chen2023diversevul}, it directly inherits its documented $5-15\%$ label noise alongside its inherent class imbalance across vulnerability categories.

\vspace{2pt}
\noindent
(ii) \textit{Model selection and adaptation}. The validity of the results may also be affected by the choice of learning models \citep{hui2024qwen25codertechnicalreport,guo2024deepseekcoderlargelanguagemodel,feng2020codebert,guo2021graphcodebertpretrainingcoderepresentations,radford2019gpt2,face2023codeparrot,wang2023codet5} and their associated adaptation configurations. While smaller encoder-based and sequence-to-sequence baselines underwent full fine-tuning, computational resource constraints needed adapting larger generative LLMs, e.g., Qwen2.5-Coder and DeepSeek-Coder, via Low-Rank Adaptation (LoRA) over 3 epochs. While the selected models and fine-tuning regimes reflect widely adopted state-of-the-art approaches, this architectural and adaptation heterogeneity introduces a potential confounding factor, as more extensive hyperparameter optimization could influence model capacity.

\vspace{2pt}
\noindent
(iii) \textit{Metrics}. To evaluate the performance of neural decompilation, we selected output similarity metrics such as BLEU, ES, and ROUGE-L. This choice may threaten construct validity, as such metrics may not capture the correctness of the decompiled code. Nevertheless, output similarity metrics represent the most popular practice for evaluating neural machine translation, both for code-to-code \citep{hosseini2022beyond, katz2018using}, and NL-to-code translation tasks \citep{liguori2023evaluates}.

\vspace{2pt}
\noindent
(iv) \textit{OOD Vulnerable Sample Size}. The limited absolute number of vulnerable samples ($29$ instances, reflecting a realistic $2.7\% \approx 3\%$ prevalence across $1,077$ functions) in our OOD test set introduces high statistical variance. A bootstrap analysis yields wide $95\%$ confidence intervals for both Vul-Recall ($95\%$ CI $[0.12, 0.44]$) and Vul-F1 ($95\%$ CI $[0.07, 0.29]$). Consequently, while this setup preserves operational realism, individual prediction shifts under extreme imbalance significantly affect point estimates and should be considered when interpreting OOD degradation.
\section{Conclusion}
\label{sec:conclusion}

In this work, we introduced VulDec, a novel large-scale dataset that aligns C/C++ vulnerability annotations with their corresponding assembly representations, together with an end-to-end framework that leverages large language model–based neural decompilation for binary vulnerability detection. By translating binary executables into source-level pseudo-code and treating both representations as textual artifacts, our approach demonstrates how neural decompilation can serve as an effective abstraction for downstream cybersecurity tasks. Our primary objective was to advance research in black-box binary vulnerability detection by enabling realistic, data-driven studies and highlighting the potential of neural decompilation as a unifying bridge between low-level binaries and source-level semantics. The experimental evaluation of out-of-distribution, real-world binaries highlights the effectiveness and limitations of the proposed approach.

\textit{Future work} will focus on enhancing the robustness of neural detectors against out-of-distribution samples, mitigating hallucinations in neural decompilers to ensure reliability in real-world scenarios, and incorporating semantic constraints and data-flow representations to enforce behavioral correctness in binary analysis.
\section*{Acknowledgments}
\label{sec:Ack}
This work was supported by the European Union – NextGenerationEU under the PNRR (Piano Nazionale di Ripresa e Resilienza), Mission 4, Component 2, Investment 'I.3.4 Borse PNRR transizioni digitali e ambientali', through Ministerial Decree no. 118 (02-03-2023). The authors would like to express their gratitude to the former student, \textit{Marco Cimmino}, for his valuable contributions to the experimental evaluation of this work.
\begin{appendices}
%\newpage
\section{Additional Details on VulDec Dataset}\label{app:vuldec}

Table~\ref{tab:ProjectFunctionCountsVersions} reports the number of functions per project in the VulDec dataset for both ARM and x86 architectures, along with the corresponding project versions. The distribution of functions highlights the predominance of large-scale projects such as Linux, which contribute the majority of samples, while smaller projects provide fewer functions. All projects and their versions are identical across the two architectures, ensuring comparability between the ARM and x86 subsets. This selection of widely used open-source projects covers diverse application domains and software types, providing a representative and challenging benchmark for function-level vulnerability detection. 

\begin{table}[htbp]
\caption{Project versions and number of functions per architecture}
%\begin{center}
%\resizebox{\columnwidth}{!}{%
\begin{tabular}{l|ccc}
\toprule
\textbf{Project} & \textbf{Version} & \textbf{ARM \#functions} & \textbf{x86 \#functions} \\
\midrule
linux & 5.4.0 & 18,965 & 19,979 \\
qemu & 4.2.0 & 580 & 2,451 \\
openssl & 1.1.1f & 1,303 & 1,311 \\
vim & 8.1.2269 & 857 & 843 \\
mysql-server & 8.0.40 & 147 & 145 \\
samba & 4.15.13 & 458 & 492 \\
libxml2 & 2.9.10 & 1,000 & 1,001 \\
FFmpeg & 4.2.7 & 910 & 911 \\
qpdf & 9.1.1 & 92 & 92 \\
wireshark & 3.2.3 & 150 & 150 \\
git & 2.21.1 & 541 & 541 \\
redis & 5.0.7 & 171 & 171 \\
curl & 7.68.0 & 223 & 222 \\
systemd & v245.4 & 670 & 669 \\
libvirt & v6.0.0 & 172 & 172 \\
bash & 5.0 & 118 & 118 \\
gnupg & 2.2.19 & 367 & 365 \\
mono & 6.8.0.105 & 144 & 146 \\
nautilus & 3.36.3 & 326 & 326 \\
OpenSC & 0.20.0 & 460 & 462 \\
libarchive & v3.4.0 & 275 & 275 \\
krb5 & 1.17-final & 308 & 308 \\
flatpak & 1.6.5 & 168 & 169 \\
bluez & 5.53 & 21 & 21 \\
\bottomrule
\end{tabular}
%}\end{center}
\label{tab:ProjectFunctionCountsVersions}
\end{table}

The dataset exhibits a skewed distribution dominated by large-scale systems, where the \textit{linux} kernel alone contributes over 60\% of the total functions. While this guaranties an exceptional representation of complex, enterprise-grade software patterns, it may limit the model's immediate generalizability to smaller projects, embedded systems, and IoT firmware. As a consequence, the learned features may fail to generalize effectively to the constrained coding conventions, lightweight libraries, and specific architectural optimizations typical of embedded and IoT environments.

\section{Experimental Setup Details}\label{app:setup}

Table \ref{tab:dataset_stats} summarizes the main statistics of the filtered and split x86 and ARM datasets used for neural decompilation, including the distribution across the training, development, and test sets. It also reports token-level statistics for both the input Assembly code and the output C code, showing that Assembly sequences are, on average, longer than their C counterparts. For completeness, the table additionally presents the statistics of the unbalanced dataset adopted for vulnerability detection.
\begin{table*}[h]
\centering
\caption{Unified statistics for neural decompilation (x86/ARM) and vulnerability detection preprocessed datasets}
\label{tab:dataset_stats}
\small
\begin{tabular}{p{2cm}p{3cm}p{2cm}p{2cm}p{2cm}}
\toprule
\textbf{Category} & \textbf{Feature} & \textbf{x86} $\rightarrow$ \textbf{C} & \textbf{ARM} $\rightarrow$ \textbf{C} & \textbf{C} $\rightarrow$ \textbf{target} \\
\midrule
\multirow{4}{*}{Split} & Total Examples & 15,624 & 12,300 & 213,661  \\
 & Train & 14,061 & 11,070 & 192,295$^{*}$\\
 & Dev & 781 & 615 &  10,683\\
 & Test & 782 & 615 & 10,683\\
\midrule
\multirow{3}{*}{Assembly} & Min Tokens & 9 & 8 & -- \\
 & Max Tokens & 512 & 512 & -- \\
 & Avg. Tokens & 227.35 & 250.87 & -- \\
\midrule
\multirow{3}{*}{C Source} & Min Tokens & 10 & 11 & 20 \\
 & Max Tokens & 256 & 256 & 256 \\
 & Avg. Tokens & 100.26 & 95.24 & 99.45 \\
 \midrule
\multirow{2}{*}{Classes} & Safe & -- & -- & 204,664 \\
 & Vulnerable & -- & -- & 8,997
 \\
\bottomrule
\multicolumn{5}{l}{\small $^{*}$ Train samples shown for the raw configuration; others (US) can be derived accordingly}
\end{tabular}
\end{table*}
\begin{table*}[htbp]
\centering
\caption{Training configuration for neural decompilation and vulnerability detection}
\label{tab:training_configs}
\small
\begin{tabular}{lcc}
\toprule
\textbf{Parameter} & \textbf{Neural Decompilation} & \textbf{Vulnerability Detection} \\
\midrule
Optimizer & AdamW & AdamW \\
Learning Rate & $5 \times 10^{-5}$ & $2 \times 10^{-5}$ \\
Epsilon (AdamW) & $1 \times 10^{-8}$ & $1 \times 10^{-8}$ \\
Weight Decay & 0 & 0.01 \\
Gradient Clipping & max\_grad\_norm = 1.0 & -- \\
Scheduler & 1000 warm-up steps & Linear decay \\
Max Source Length & 512 tokens & 256 tokens \\
Max Target Length & 256 tokens & Binary output (0/1) \\
Split & 90/5/5 & 90/5/5 \\
\bottomrule
\end{tabular}
\end{table*}

Table \ref{tab:training_configs} summarizes the key parameters used in the VulDec pipeline. The workflow follows a progressively decreasing input/output size: the first stage processes assembly code with a maximum length of 512 tokens and generates C/C++ code limited to 256 tokens, while the final output encodes binary vulnerability information (0/1). This design reflects the hierarchical nature of the task, moving from low-level instructions to high-level program semantics and finally to a compact, security-relevant representation. The dataset splits are chosen to balance sufficient training data with effective evaluation, using 90/5/5 for both tasks, ensuring model generalization and robust testing. A block of parameters, i.e., decision threshold, batch size, gradient accumulation, number of epochs, and beam size, was adapted for each model and each task to balance computational constraints, training time, and performance expectations, reflecting differences in model complexity and architecture. Training budgets were scaled by architecture and pre-training capacity. Encoder-decoder models (CodeT5, PLBART) required longer schedules (up to $10$ epochs) for cross-attention alignment. Conversely, large decoder-only LLMs (Qwen, DeepSeek) were capped at $3$ epochs to prevent overfitting. Smaller decoder-only models (CodeGPT, CodeParrot) were allocated an intermediate budget ($5$--$8$ epochs) to compensate for their lower capacity, while encoder-only models received minimal epochs due to their generative unsuitability. 

\section{Class-Imbalance Strategy Optimization across Model Architectures}\label{app:strategy}

\begin{table*}[htbp]
\centering
\caption{Class-imbalance mitigation strategies extended to all model architectures on validation set (3\% vulnerability prevalence). CW: Class Weighting; US: Under-Sampling; SLM: Small Language Model; LLM: Large Language Model}
\label{tab:full_strategy}

\resizebox{\textwidth}{!}{
\begin{tabular}{cl|cccc}
\toprule
\textbf{Architecture (Model)}
& \textbf{Strategy}
& \textbf{Vul-Prec}
& \textbf{Vul-Recall}
& \textbf{Vul-F1}
& \textbf{F1-score}
\\

\midrule
\multirow{4}{*}{\shortstack{Encoder Only SLM\\(CodeBERT)}}
& US 50-50
& 0.070
& 0.707
& 0.128
& 0.466
\\
& US 70-30
& 0.127
& 0.427
& 0.196
& 0.567
\\
& US 80-20
& 0.167
& 0.336
& 0.223
& 0.581
\\

\cmidrule{2-6}
& \textbf{US 80-20 + CW}
& 0.171
& 0.415
& \textbf{0.243}
& 0.591
\\

\midrule
\multirow{4}{*}{\shortstack{Encoder-Decoder SLM\\(PLBART)}}

& US 50-50
& 0.072
& 0.725
& 0.131
& 0.469
\\
& US 70-30
& 0.131
& 0.450
& 0.203
& 0.572
\\
& US 80-20
& 0.170
& 0.360
& 0.231
& 0.587
\\
\cmidrule{2-6}
& \textbf{US 80-20 + CW}
& 0.178
& 0.459
& \textbf{0.257}
& 0.605
\\

\midrule
\multirow{4}{*}{\shortstack{Decoder Only SLM\\(CodeGPT)}}
& US 50-50
& 0.068
& 0.685
& 0.124
& 0.458
\\
& US 70-30
& 0.120
& 0.420
& 0.187
& 0.559
\\
& US 80-20
& 0.155
& 0.340
& 0.213
& 0.576
\\
\cmidrule{2-6}

& \textbf{US 80-20 + CW}
& 0.162
& 0.420
& \textbf{0.234}
& 0.586
\\

\midrule
\multirow{3}{*}{\shortstack{Decoder Only LLM\\(Qwen2.5-Coder)}}
& \textbf{US 50-50}
& 0.175
& 0.320
& \textbf{0.226}
& 0.588
\\
& US 70-30
& 0.182
& 0.225
& 0.201
& 0.570
\\
& US 80-20
& 0.158
& 0.219
& 0.183
& 0.575
\\

\midrule
\multirow{3}{*}{\shortstack{Decoder Only LLM\\(DeepSeek-Coder)}}
& \textbf{US 50-50}
& 0.179
& 0.272
& \textbf{0.221}
& 0.585
\\
& US 70-30
& 0.162
& 0.205
& 0.181
& 0.571
\\
& US 80-20
& 0.140
& 0.165
& 0.151
& 0.560
\\

\bottomrule
\end{tabular}
}
\end{table*}

In order to adopt the optimal mitigation strategy under extreme class imbalance, we evaluated different under-sampling ratios on the validation set across representative model architectures, i.e., CodeBERT for Encoder-Only, PLBART for Encoder-Decoder, and CodeGPT for small Decoder-Only. Additionally, we explicitly performed this optimization for both modern decoder-only LLMs (Qwen2.5-Coder and DeepSeek-Coder) to rigorously verify their specific training dynamics.

As shown in Table \ref{tab:full_strategy}, for standard encoder-based backbones as well as small decoder-only models, the combination of \textbf{US 80-20 + CW} consistently achieves the best balance between precision and recall, yielding the highest Vul-F1 and Macro F1 scores. Conversely, for modern decoder-only LLMs, the \textbf{US 50-50} sampling ratio proves to be the most effective strategy. Since class weights are defined as inversely proportional to the training class distribution, applying CW to a 50-50 balanced dataset results in a weight ratio of 1:1, making loss weighting mathematically redundant. Therefore, class balance was controlled entirely via dataset resampling without additional loss weighting.

\section{CWE Macrocategory Mapping}\label{app:map}

Table \ref{tab:map} presents the costum mapping between the defined macro-categories and the corresponding CWE identifiers. Each macro-category groups vulnerabilities sharing common security characteristics

\begin{table}[!htbp]
\caption{Mapping between macrocategories and associated CWE identifiers}
\label{tab:map}
\centering
\begin{tabular}{p{5cm}|p{7cm}}
\toprule
\textbf{Macro-Category} & \textbf{Included CWEs} \\
\midrule

Taint Propagation Issues &
CWE-20, CWE-74, CWE-77, CWE-79, CWE-89, CWE-116 \\
\midrule

Insecure Configuration Issues &
CWE-16, CWE-611, CWE-276, CWE-732, CWE-933 \\
\midrule

Data Protection Issues &
CWE-200, CWE-201, CWE-312, CWE-319, CWE-522 \\
\midrule

Validation and Authorization Issues &
CWE-703, CWE-285, CWE-862, CWE-863, CWE-306, CWE-807 \\
\midrule

Memory Management and Concurrency &
CWE-119, CWE-120, CWE-125, CWE-416, CWE-476, CWE-362, CWE-401 \\
\midrule

Uncategorized (Remaining CWEs) &
This category aggregates all safe and vulnerable functions for which the CWE column is empty. \\
\bottomrule

\end{tabular}
\end{table}

\section{Decompiled Code: Qualitative Analysis}\label{app:decompiled}

\begin{table*}[!htbp]
\caption{Qualitative analysis of VulDec neural decompilation outputs across different performance tiers sorted by Edit Similarity (ES)}
\label{tab:qualitative}
%\footnotesize
\centering
\hbadness=10000
\resizebox{0.95\textwidth}{!}{
\begin{tabular}{
 >{\arraybackslash}m{8.5cm} 
 >{\arraybackslash}m{8.5cm}}
\toprule
\textbf{Ground Truth} & \textbf{VulDec Decompiled} \\ \midrule

% ==================== BLOCCO 1: TIER 1 (ES: 95.21%) ====================
%\multicolumn{2}{l}{\textit{\textbf{Tier 1: Perfect Structural Match \& Contextual Renaming}}} \\ \addlinespace[0.1cm]
\texttt{store\_tabletJitterDelay(\textcolor{blue}{struct} device *dev,} & \texttt{store\_tabletJitterDelay(\textcolor{blue}{struct} device *dev,} \\
\hspace{0.4cm}\texttt{\textcolor{blue}{struct} device\_attribute *attr, \textcolor{blue}{const char} *buf,} & \hspace{0.4cm}\texttt{\textcolor{blue}{struct} device\_attribute *attr, \textcolor{blue}{const char} *buf,} \\
\hspace{0.4cm}\texttt{\textcolor{blue}{size\_t} count)} & \hspace{0.4cm}\texttt{\textcolor{blue}{size\_t} count)} \\
\texttt{\{} & \texttt{\{} \\
\hspace{0.4cm}\texttt{\textcolor{blue}{struct} aiptek *aiptek = dev\_get\_drvdata(dev);} & \hspace{0.4cm}\texttt{\textcolor{blue}{struct} aiptek *aiptek = dev\_get\_drvdata(dev);} \\
\hspace{0.4cm}\texttt{\textcolor{blue}{int} err, \textcolor{blue}{j};} & \hspace{0.4cm}\texttt{\textcolor{blue}{int} err, \textcolor{blue}{delay};} \\
\addlinespace[0.05cm]
\hspace{0.4cm}\texttt{err = kstrtoint(buf, 10, \&\textcolor{blue}{j});} & \hspace{0.4cm}\texttt{err = kstrtoint(buf, 10, \&\textcolor{blue}{delay});} \\
\hspace{0.4cm}\texttt{\textcolor{blue}{if} (err) \textcolor{blue}{return} err;} & \hspace{0.4cm}\texttt{\textcolor{blue}{if} (err) \textcolor{blue}{return} err;} \\
\addlinespace[0.05cm]
\hspace{0.4cm}\texttt{aiptek->newSetting.jitterDelay = \textcolor{blue}{j};} & \hspace{0.4cm}\texttt{aiptek->newSetting.jitterDelay = \textcolor{blue}{delay};} \\
\hspace{0.4cm}\texttt{\textcolor{blue}{return} count;} & \hspace{0.4cm}\texttt{\textcolor{blue}{return} count;} \\
\texttt{\}} & \texttt{\}} \\ \addlinespace[0.1cm]
\multicolumn{2}{c}{\textbf{Edit Similarity (ES): 95.21\%}} \\ \midrule

% ==================== BLOCCO 2: TIER 2 (ES: 93.15%) ====================
%\multicolumn{2}{l}{\textit{\textbf{Tier 2: Syntactic Shift with Functional Equivalence}}} \\ \addlinespace[0.1cm]
\texttt{s32 i2c\_smbus\_read\_byte\_data(\textcolor{blue}{const struct} i2c\_client} & \texttt{s8 i2c\_smbus\_read\_byte\_data(\textcolor{blue}{const struct} i2c\_client} \\
\hspace{0.4cm}\texttt{*client, u8 command)} & \hspace{0.4cm}\texttt{*client, u8 command)} \\
\texttt{\{} & \texttt{\{} \\
\hspace{0.4cm}\texttt{union i2c\_smbus\_data data;} & \hspace{0.4cm}\texttt{union i2c\_smbus\_data data;} \\
\hspace{0.4cm}\texttt{\textcolor{blue}{int} status;} & \hspace{0.4cm}\texttt{\textcolor{blue}{int} status;} \\
\addlinespace[0.05cm]
\hspace{0.4cm}\texttt{status = i2c\_smbus\_xfer(client->adapter, client->addr, client->flags, I2C\_SMBUS\_READ, command, I2C\_SMBUS\_BYTE\_DATA, \&data);} & \hspace{0.4cm}\texttt{status = i2c\_smbus\_xfer(client->adapter, client->addr, client->flags, I2C\_SMBUS\_READ, command, I2C\_SMBUS\_BYTE\_DATA, \&data);} \\
\hspace{0.4cm}\texttt{\textcolor{blue}{return} (status < 0) ? status : data.byte;} & \hspace{0.4cm}\texttt{\textcolor{blue}{if} (status < 0) \textcolor{blue}{return} status;} \\
 & \hspace{0.4cm}\texttt{\textcolor{blue}{return} data.byte;} \\
\texttt{\}} & \texttt{\}} \\ \addlinespace[0.1cm]
\multicolumn{2}{c}{\textbf{Edit Similarity (ES): 93.15\%}} \\ \midrule

% ==================== BLOCCO 3: TIER 3 (ES: 58.15%) ====================
%\multicolumn{2}{l}{\textit{\textbf{Tier 3: Intermediate Tier Example (Semantic/Type Shift)}}} \\ \addlinespace[0.1cm]
\texttt{\textcolor{blue}{static int} cmp\_extents\_reverse(\textcolor{blue}{const void} *a, \textcolor{blue}{const void} *b) \{} & \texttt{\textcolor{blue}{static int} cmp\_extents\_reverse(\textcolor{blue}{const void} *a, \textcolor{blue}{const void} *b) \{} \\
\hspace{0.4cm}\texttt{\textcolor{blue}{const struct} uid\_gid\_extent *e1 = a;} & \hspace{0.4cm}\texttt{\textcolor{blue}{const struct} extent\_map *em\_a = *(\textcolor{blue}{const struct} extent\_map **)a;} \\
\hspace{0.4cm}\texttt{\textcolor{blue}{const struct} uid\_gid\_extent *e2 = b;} & \hspace{0.4cm}\texttt{\textcolor{blue}{const struct} extent\_map *em\_b = *(\textcolor{blue}{const struct} extent\_map **)b;} \\
\addlinespace[0.05cm]
\hspace{0.4cm}\texttt{\textcolor{blue}{if} (e1->lower\_first < e2->lower\_first)} & \hspace{0.4cm}\texttt{\textcolor{blue}{if} (em\_a->start < em\_b->start)} \\
\hspace{0.8cm}\texttt{\textcolor{blue}{return} -1;} & \hspace{0.8cm}\texttt{\textcolor{blue}{return} 1;} \\
\hspace{0.4cm}\texttt{\textcolor{blue}{if} (e1->lower\_first > e2->lower\_first)} & \hspace{0.4cm}\texttt{\textcolor{blue}{if} (em\_a->start > em\_b->start)} \\
\hspace{0.8cm}\texttt{\textcolor{blue}{return} 1;} & \hspace{0.8cm}\texttt{\textcolor{blue}{return} -1;} \\
\hspace{0.4cm}\texttt{\textcolor{blue}{return} 0;} & \hspace{0.4cm}\texttt{\textcolor{blue}{return} 0;} \\
\texttt{\}} & \texttt{\}} \\ \addlinespace[0.1cm] 
\multicolumn{2}{c}{\textbf{Edit Similarity (ES): 58.15\%}} \\ \midrule

% ==================== BLOCCO 4: TIER 4 (ES: 17.62%) ====================
%\multicolumn{2}{l}{\textit{\textbf{Tier 4: Worst Tier Example (Infinite Repetition Loop)}}} \\ \addlinespace[0.1cm]
\texttt{xmlRegStatePtr xmlRegNewState( xmlRegParserCtxtPtr ctxt) \{} & \texttt{xmlRegStatePtr xmlRegNewState( xmlRegParserPtr parser) \{} \\
\hspace{0.4cm}\texttt{xmlRegStatePtr ret;} & \hspace{0.4cm}\texttt{xmlRegParserStatePtr state;} \\
\addlinespace[0.05cm]
\hspace{0.4cm}\texttt{ret = (xmlRegStatePtr) xmlMalloc(\textcolor{blue}{sizeof}(xmlRegState));} & \hspace{0.4cm}\texttt{state = (xmlRegParserStatePtr) xmlMalloc(\textcolor{blue}{sizeof}(xmlRegParserState));} \\
\hspace{0.4cm}\texttt{\textcolor{blue}{if} (ret == NULL) \{} & \hspace{0.4cm}\texttt{\textcolor{blue}{if} (state == NULL) \textcolor{blue}{return}(NULL);} \\
\hspace{0.8cm}\texttt{xmlRegexpErrMemory(ctxt, "allocating state");} & \hspace{0.4cm}\texttt{state->state = XML\_REG\_STATE\_INITIAL;} \\
\hspace{0.8cm}\texttt{\textcolor{blue}{return}(NULL);} & \hspace{0.4cm}\texttt{state->error = 0;} \\
\hspace{0.4cm}\texttt{\}} & \hspace{0.4cm}\texttt{state->errorLine = 0;} \\
\hspace{0.4cm}\texttt{memset(ret, 0, sizeof(xmlRegState));} & \hspace{0.4cm}\texttt{state->errorColumn = 0;} \\
\hspace{0.4cm}\texttt{ret->type = XML\_REGEXP\_TRANS\_STATE;} & \hspace{0.4cm}\texttt{state->errorOffset = 0;} \\
\hspace{0.4cm}\texttt{ret->mark = XML\_REGEXP\_MARK\_NORMAL;} & \hspace{0.4cm}\texttt{\dots} \\
\hspace{0.4cm}\texttt{\textcolor{blue}{return}(ret);} & \hspace{0.4cm}\texttt{\textit{/* ... truncated due to infinite repetition loop ... */}} \\
\texttt{\}} & \texttt{\}} \\ \addlinespace[0.1cm] 
\multicolumn{2}{c}{\textbf{Edit Similarity (ES): 17.62\%}} \\ \bottomrule
\end{tabular}
}
\end{table*}

The qualitative analysis reveals both strengths and limitations of the x86 neural decompiler. As shown in Table \ref{tab:qualitative}, we report four pairs of reference and decompiled functions extracted from the test set.

\vspace{2pt}
\noindent
\textbf{Tier 1: Perfect Structural Match and Contextual Renaming}.
This tier represents the ideal success scenario for our neural framework. As shown in the \texttt{store\_tabletJitterDelay} routine from the Linux Kernel input subsystem, the generated control-flow graph and syntactic blocks mirror the ground truth perfectly. The marginal reduction in the ES score (95.21\%) is exclusively caused by a variable renaming effect, where the original single-character identifier \texttt{j} is translated into \texttt{delay}. Far from being a flaw, this behavior highlights the model's ability to extract semantic context from surrounding tokens (e.g., the function name and the structure field \texttt{jitterDelay}).

\vspace{2pt}
\noindent
\textbf{Tier 2: Syntactic Shift with Functional Equivalence}.
The second tier highlights cases where the model demonstrates a deep semantic understanding derived from its pre-training phase. In the \texttt{i2c\_smbus\_read\_byte\_data} function, the ground truth implements a compact inline instruction: \texttt{return (status < 0) ? status : data.byte;}. VulDec decompiler successfully interprets the low-level conditional branch layout but decomposes it into an explicit, structurally expanded code path: \texttt{if (status < 0) return status; return data.byte;}. Although this syntactic variance induces a minor drop in token-level similarity (93.15\%), the decompiled function remains 100\% functionally equivalent.

\vspace{2pt}
\noindent
\textbf{Tier 3: Semantic and Type Shift}.
The intermediate tier (ES: 58.15\%) uncovers critical insights regarding prompt-induced biases in generative models. In the \texttt{cmp\_extents\_reverse} example, the original code performs a standard three-way ascending comparison. However, the presence of the word \texttt{"reverse"} within the function identifier biases the autoregressive decoding, prompting the LLM to assume a descending sort order. Consequently, the model generates a specular return logic, swapping the \texttt{1} and \texttt{-1} outputs. Additionally, it experiences a type misalignment, injecting an extraneous pointer indirection (\texttt{*(const struct extent\_map **)a}) and swapping the \texttt{uid\_gid\_extent} type with \texttt{extent\_map}. Despite the semantic inversion, the underlying conditional structure (\texttt{if-if-return}) is accurately preserved.

\vspace{2pt}
\noindent
\textbf{Tier 4: Catastrophic Hallucination and Loop}.
This final tier (ES: 17.62\%) embodies the catastrophic failure modes typical of autoregressive language models when handling unaligned binary syntax or hard code structures. Our extensive empirical evaluations indicate that severe hallucinations generally manifest as one of four cases: 
\begin{enumerate}
    \item premature decoding termination immediately after generating the function signature, omitting the function body entirely;
    \item generation of a correct code block followed by arbitrary syntactic noise, unaligned code paths, or the same block repeated in a loop;
    \item the fluent generation of a structurally correct but completely unrelated function;
    \item a severe collapse of the model's self-attention stopping criteria, leading to an infinite repetition loop on a single line of code.
\end{enumerate}
The \texttt{xmlRegNewState} routine from \texttt{libxml2} showcases this latter attention collapse pattern. After hallucinating a non-existent error-tracking structure (\texttt{errorLine}, \texttt{errorColumn}, etc.), the decoder becomes trapped in an infinite loop, continuously reiterating identical allocation sequences until truncated by the maximum token generation boundary, leaving an incomplete and non-compilable artifact.

\end{appendices}

\bibliography{sn-bibliography}% common bib file


\end{document}